\documentclass{IEEEtran}
\usepackage{cite}
\usepackage{amsmath,amssymb,amsfonts}
\usepackage{algorithmic}
\usepackage[caption=false,font=footnotesize]{subfig}
\usepackage{graphicx}
\usepackage{placeins}
\usepackage{textcomp}
\usepackage{booktabs}
\usepackage{xcolor}
\usepackage{threeparttable}
\usepackage{siunitx}
\usepackage{tabularx}  
\usepackage{makecell}   
\usepackage{balance}  
\usepackage{dblfloatfix}
\usepackage{balance}
\usepackage[normalem]{ulem}
\usepackage{hyperref}
\newcolumntype{Y}{>{\raggedright\arraybackslash}X}
\renewcommand\arraystretch{1.12} 
\def\BibTeX{{\rm B\kern-.05em{\sc i\kern-.025em b}\kern-.08em
    T\kern-.1667em\lower.7ex\hbox{E}\kern-.125emX}}
\begin{document}
\title{Harnessing Selective State Space Models to Enhance Semianalytical Design of Fabrication- Ready Multilayered Huygens' Metasurfaces:\\Part II – Generative Inverse Design (MetaMamba)}
\author{Natanel Nissan,~\IEEEmembership{Student Member, IEEE}, Sherman W. Marcus,~\IEEEmembership{Life Member, IEEE}, \\Dan Raviv,~\IEEEmembership{Member, IEEE}, Raja Giryes,~\IEEEmembership{Senior Member, IEEE}, and Ariel Epstein,~\IEEEmembership{Senior Member, IEEE}%
\thanks{The work was supported by the Israel Innovation Authority through its
Metamaterials consortium.}%
\thanks{N. Nissan, D. Raviv, and R. Giryes are with the School of Electrical and Computer Engineering, Tel Aviv University, Tel Aviv 69978, Israel (e-mails: natanel.nissan@gmail.com, raviv.dan@gmail.com, raja@tauex.tau.ac.il).}%
\thanks{S. W. Marcus and A. Epstein are with the Andrew and Erna Viterbi Faculty of Electrical and Computer Engineering, Technion – Israel Institute of Technology, Haifa 3200003, Israel (e-mails: shermanm@technion.ac.il, epsteina@ee.technion.ac.il).}%
\thanks{The code developed and utilized in Part I and Part II of this two-part compilation is publicly available at https://github.com/nati223/metamamba.}
}

\maketitle

\begin{abstract}
We present a generative framework for inverse design of five-layer transmissive Huygens' metasurfaces (HMSs), addressing a longstanding challenge in achieving full-phase, high-efficiency unit cell designs with minimal full-wave simulations. The key to achieving this is our reliance on the field-based semianalytical (SA) scheme developed in Part I of this paper, which allows rapid and highly effective synthesis of such multilayer composites, however with limited accuracy. To overcome the prohibitive data demands of traditional pipelines, we employ Mamba, a selective state space model well suited for long-range sequence modeling as the backbone of our learning framework. A bidirectional Mamba (Bi-Mamba) forward surrogate is first trained on SA-generated data and subsequently fine-tuned with full-wave CST samples. An ablation over a 1080-sample CST pool shows that as few as 270 full-wave calibration samples suffice to reach near--CST-level agreement at a fraction of the simulation cost. An autoregressive Mamba inverse generator is subsequently trained on surrogate-augmented data, treating unit-cell synthesis as a sequential generation task. The resulting one-to-many generative model produces diverse unit cell geometries conditioned on target scattering responses. It achieves CST-validated designs with field transmission magnitude 0.9 across the full 0–$2\pi$ phase range at 20~GHz. Moreover, a CST-calibrated surrogate trained to accurately predict frequency responses (18–22~GHz) enables functional post-selection of inverse generated designs.
Together, the hybrid SA–generative methodology in this two-part compilation establishes a scalable and data-efficient solution for multilayer HMS synthesis, with natural 
extensions toward broadband, oblique-incidence, and higher-dimensional electromagnetic inverse-design problems.
\end{abstract}


\section{Introduction}
\label{sec:introduction}
Electromagnetic metasurfaces - low-profile arrays of subwavelength polarizable particles (meta-atoms) - provide compact, planar means to manipulate wavefronts, enabling versatile control of amplitude, phase, and polarization~\cite{Holloway2005TEMC,Glybovski2016PhysRep}. Within this class, Huygens' metasurfaces (HMSs) are particularly attractive, as their co-located electric and magnetic polarizabilities enable a wide range of power-conserving field transformations with minimal reflection~\cite{Pfeiffer2013PRL, Monticone2013Full, Selvanayagam2013Discontinuous, Epstein2016JOSAB,Epstein2016TAP}. At the macroscopic level, HMS transformations can be rigorously described by the generalized sheet transition conditions (GSTCs), which relate discontinuities in the electromagnetic fields to surface susceptibility tensors~\cite{Kuester2003TAP, Tretyakov2003Analytical, Tretyakov2015TAP, Achouri2015TAP}. As discussed in Part~I~\cite{Marcus2026LAYERS}, for common beamforming applications (e.g., anomalous refraction plates or focusing lenses), GSTC-based designs typically specify local responses corresponding to reflectionless meta-atoms that impose prescribed phase shifts when illuminated by a normally incident plane wave. However, the inverse design of such transmissive HMS unit cells (microscopic design~\cite{Epstein2016JOSAB}), achieving near-unity transmission while covering the full \(2\pi\) phase range in realistic subwavelength platforms, continues to be particularly non-trivial. While GSTCs prescribe the required surface response, translating them into manufacturable geometries remains challenging, missing a systematic, reliable, and time-effective methodology that could produce realistic constructs that satisfy the specified strict scattering constraints.

At microwave frequencies, the most natural way to realize this required co-located electric and magnetic response is via a conducting loop and wire combination defined on printed circuit boards (PCBs) \cite{Pfeiffer2013PRL, Chen2020Omega}. Nonetheless, the latter have to be cut and rearranged manually, resulting in a complex multi-board formation that may be unsuitable for many applications. Alternatively, it was shown that implementation using symmetric cascades of thin electrically polarizable sheets (impedance sheets) may provide sufficient degrees of freedom to support excitation of symmetric and antisymmetric modes, effectively emulating the necessary electric and magnetic surface polarizabilities for Huygens' meta-atoms \cite{Monticone2013Full,Pfeiffer2013Millimeter}. This structure is fully compatible with simple and standard vialess multilayer PCB configuration, forming a highly appealing realization platform \cite{Epstein2016JOSAB, Chen2023MAM}. Nevertheless, multiple stacked layers are typically required for high-efficiency operation~\cite{Pfeiffer2014Bianisotropic,AbdoSanchez2019A}, while brute-force optimization becomes intractable as layer count grows.
Transmission-line based models considering multiple reflections between the homogenized impedance sheets \cite{Monticone2013Full,Pfeiffer2013Millimeter,Pfeiffer2014Bianisotropic,Epstein2016TAP,AbdoSanchez2019A} offer instantaneous $S$-parameter predictions and may shed light on important physical interplays, but suffer significant inaccuracies due to neglected interlayer near field coupling effects, becoming more pronounced for the small interlayer spacing in common metasurfaces (MSs). Thus, in practically all cases, ultimate fine-tuning via full-wave optimizers is necessary to obtain a working design \cite{Pfeiffer2014Bianisotropic, Chen2018Theory, Lavigne2018Susceptibility}. Attempts to incorporate mutual coupling through dedicated components in extended circuit models had some success in this sense, but require iterative adjustment of their values (or prior knowledge on coupling properties), reducing the versatility and attractiveness of such approaches \cite{Xu2018ATechnique,Olk2019Accurate}.

In Part~I of this two part paper, we have proposed and systematically formulated a field-based semianalytical (SA) methodology that can rigorously account also for these intricate near field effects (manifested by the evanescent wave spectrum) for multilayer loaded-wire-based meta-atom configurations \cite{Tretyakov2003Analytical, Rabinovich2020Arbitrary}, facilitating diverse fabrication-ready MS designs with improved fidelity~\cite{levy2019rigorous, levy2019synthesis, Killamsetty2021, Kuznetsov2024Efficient,MateosRuiz2025TAP}. However, despite succeeding in devising practical and power-efficient dual-polarized frequency-diverse Huygens' meta-atom PCB layouts \emph{on demand}, some of the underlying model assumptions incur minor inaccuracies, which accumulate when considering a large number of degrees of freedom (large number of layers in the PCB stack) and might lead to deterioration of the overall meta-atom response in certain cases. Mainly, while the model properly considers near-field coupling stemming from the dominant dipole moment of the induced current on each of the loaded wires, higher-order multipoles are not taken into account, which may yield deviation in the prediction of the scattering parameters. Therefore, as laid out in Part~I\cite{Marcus2026LAYERS}, once a suitable set of Huygens' meta-atom configurations  
(copper trace specifications for each of the layers in the PCB stack) found to cover the entire $2\pi$ phase space with high efficiency is produced by the SA approach, a final verification and filtering with the aid of high-fidelity solvers (e.g., CST Microwave Studio) is conducted; only the solutions that perform well according to the simulation results (considered as "ground truth") are kept, forming a lookup table (LUT) that can then be used for the HMS macroscopic design \cite{Epstein2016JOSAB}.

Even though the relatively high fidelity of the SA model makes this design procedure highly efficient, limiting the number of numerical simulation runs to only those suitable candidate geometries output by the SA scheme - a small fraction compared to brute-force full-wave optimization - the need to verify SA predictions ultimately poses certain shortcomings. First, while the LUT effectively spans the entire $2\pi$ phase interval, in these regions where the SA model tends to yield less accurate results, the phase resolution may be limited (since only a partial set of the proposed solutions was retained after the described "filtering" procedure). Second, despite allowing rapid generation of such LUT for realistic fabrication-ready designs, it is difficult to assess whether the obtained meta-atom efficiencies are \emph{the best} one may reach with the considered physical multilayer structure for each of the considered phase shift (since the SA predicted optimum may not fully agree with full-wave "ground truth").

In parallel with physics-based acceleration methods, machine learning (ML) has emerged as a promising alternative to brute-force optimization and computationally expensive full-wave simulations for MS design~\cite{Mishra2023}. Initial efforts focused on the forward problem, employing surrogate models to map geometric parameters to electromagnetic responses~\cite{Qiu2019Deep}, thereby enabling rapid evaluations. To address the inverse design challenge, generative models were later introduced. Early studies adopted variational autoencoders (VAEs) and generative adversarial networks (GANs) for unit cell design in frequency selective surfaces (FSSs) and multilayer multi-band reflective polarizing MSs~\cite{Naseri2021TAP,naseri2022gan}. These were followed by works targeting more advanced tasks, including the inverse synthesis of transmissive MS unit cells~\cite{Wang2023GAN}, full-space quadrature-channel MS patterns~\cite{Liu2024_gan_fullspace}, and wide-angle dual-polarized FSSs using loop topologies~\cite{LyuGan25}. In parallel, diffusion models have also been explored for unit cell generation in high-efficiency transmissive MSs, achieving substantial results~\cite{Niu2023OJAP,Chen2025controlnet}.

While these generative approaches demonstrated strong potential in automating inverse MS design, they also revealed a key bottleneck: the enormous data requirement. Most works require orders of $10^4$–$10^5$ high-fidelity full-wave simulations or equivalent core-hours for training~\cite{Naseri2021TAP,naseri2022gan,Oliver2022reflect,Niu2023OJAP,Liu2024_gan_fullspace,Chen2025controlnet,Yang2025_tandem,jung2024res}. 
Recent efforts have sought to alleviate this computational burden, e.g. by employing prior knowledge-guided generative deep learning approaches for multi-objective synthesis~\cite{LyuGan25}. However, despite using physical and experiential knowledge to constrain the design space and guide network training and data acquisition, effective training still requires thousands of high-fidelity full-wave simulations \cite{LyuGan25}. 
In optics, this challenge can often be mitigated by fast approximate solvers like rigorous coupled-wave analysis (RCWA), enabling massive datasets~\cite{Yang2024_diffusion_sparse, Hen2025Diffusion}. In microwave multilayer HMS design, however, the typical reliance on full-wave solvers (e.g. CST Microwave Studio or ANSYS HFSS) makes scaling to deeper stacks particularly costly\cite{sarker2023mlantennas}.

Beyond computational limitations, existing transmissive HMS realizations also face inherent performance constraints. Achieving near-unity field transmission magnitude ($|T|$) together with full $0-2\pi$ phase ($\phi$) coverage in practical multilayer Huygens’ meta-atoms is, by itself, a highly challenging objective~\cite{Xue2020broadband,Wang2023APL,Chen2025controlnet}. To the best of our knowledge, no passive PCB-compatible configuration has demonstrated $|T| > 0.9$ across the full phase range.

In this Part II, we introduce \emph{MetaMamba}, a hybrid SA and generative framework for the inverse design of multilayer transmissive HMSs, designed to overcome these limitations. The key motivation is to combine the complementary strengths of both approaches: SA models offer near-physical accuracy and orders-of-magnitude faster runtimes than full-wave solvers, making them ideal for generating large-scale synthetic datasets; generative models, in turn, can learn rich mappings from data and achieve near-optimal inverse designs — but often require extensive datasets for training. By pretraining on SA-generated data and fine-tuning with a small number of high-fidelity full-wave simulations, MetaMamba achieves high accuracy with minimal simulation cost, effectively bridging this gap.
To capture the strong electromagnetic coupling across layers in multilayer HMSs, we formulate the design process as a sequence modeling task~\cite{Goodfellow2016}. The forward problem issues the prediction of the scattering response is inherently bidirectional, as all layers mutually interact. Conversely, the inverse problem involves auto-regressive (AR) generation~\cite{graves2014generatingsequences}: each layer is synthesized sequentially based on the target response and the previously generated layers.
This formulation naturally leads to the use of state space models (SSMs) - a class of efficient sequence models rooted in linear dynamical systems and control theory~\cite{Gu2021S4,alonso2024statespacemodelsfoundation}, where information propagates through latent hidden states that evolve in time according to learned system matrices. This formulation parallels how feedback systems or recursive filters model physical dynamics, making it particularly well suited to problems involving interdependent layers or spatial sequences. The Mamba architecture, a recent SSM variant, extends this concept by allowing the model to dynamically modulate which parts of the sequence are emphasized during state updates, achieving both global context propagation and linear-time efficiency across diverse domains~\cite{Gu2024Mamba}.
The second-generation variant, Mamba-2~\cite{Dao2024Mamba}, further stabilizes training, improves parallelism, and enhances long-sequence handling. Its ability to propagate information across stacked layers with low overhead makes it highly fit for multilayer HMS design, where strong interlayer coupling requires modeling dependencies across the entire sequence.

The MetaMamba pipeline has two main components. First, a bidirectional Mamba (Bi-Mamba) \cite{Liang2024BiMambaPlus} forward surrogate is trained in two phases: it is pretrained on 524,000 SA samples to capture broad scattering trends, and later fine-tuned with 270 high-fidelity CST simulations to closely match CST-level accuracy at minimal training and simulation time. Second, an AR-Mamba inverse generator samples layer-by-layer geometries conditioned on a desired scattering response, mimicking the structure of AR language models. This enables one-to-many generation, producing diverse and valid multilayer HMS designs.
In addition, MetaMamba extends naturally to broadband unit cell design, with a calibrated surrogate enabling efficient prediction and functional design post-selection across a range of frequencies. 
This hybrid strategy produces CST-validated, five-layer HMS unit cells with near-unity transmission and complete phase coverage, while reducing data requirements by orders of magnitude compared to full-wave-only pipelines.
Beyond the specific demonstrations provided in this two-part paper for transmissive HMS lenses, the proposed framework generalizes naturally to a broad class of electromagnetic inverse problems that can be cast as sequential generation tasks.

Taken together, Parts~I and~II establish a scalable and versatile methodology for synthesizing fabrication-ready, dual-polarized, frequency-diverse HMS designs. The computationally efficient SA scheme developed in Part~I, when coupled with the generative accuracy of the MetaMamba algorithm introduced in this work, creates a new class of hybrid physics–ML design tools. These tools combine the strengths of both worlds: physical interpretability, rapid evaluation, and near CST-level accuracy, while requiring only a minimal number of full-wave simulations for calibration. This hybrid approach provides a practical path toward on-demand inverse design of multilayer 
PCB-compatible superstrates, offering the scalability needed for deep MS stacks and large design spaces. As such, it promises to significantly accelerate the deployment of 
high-performance HMS-based beamforming and wavefront-engineering components across timely 
applications, including 5G/6G communication systems, satellite links, and next-generation 
microwave imaging platforms~\cite{Wu2016SatCom, Imani2020MicroImag, Tariq2021Mimo5G, Schlezinger2021Mimo6G, Chen2023MAM}.

\section{Method}
\label{sec:method}
\subsection{General Framework and Rationale}
\label{subsec:general_framework}
\begin{figure}[t]
    \centering
    
    \subfloat[]{%
        \includegraphics[width=0.48\linewidth]{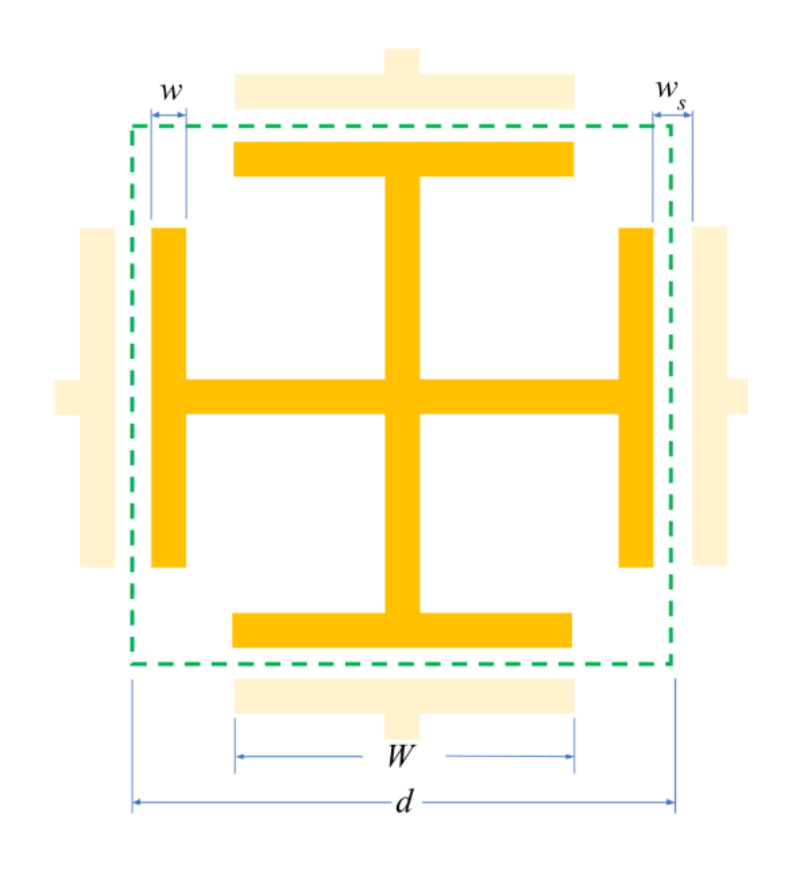}%
        \label{fig:unitcell_a}
    }
    \hfill 
    \subfloat[]{%
        \includegraphics[width=0.48\linewidth]{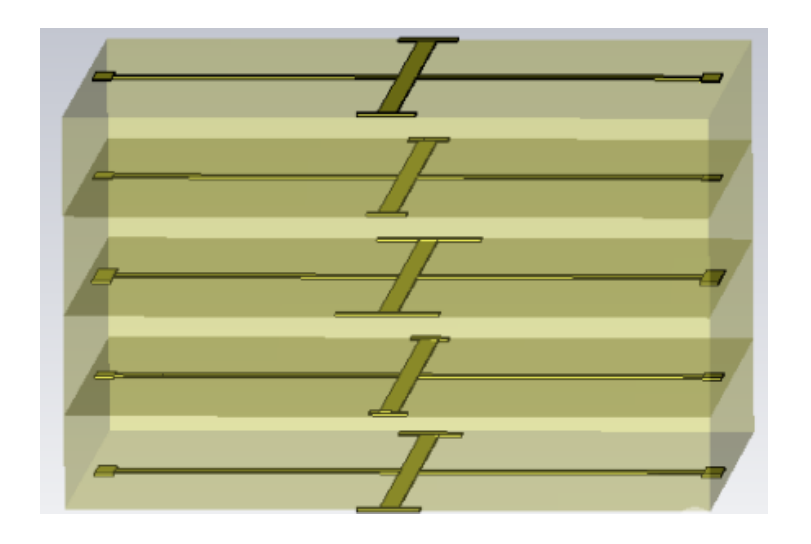}%
        \label{fig:unitcell_b}
    }

    \caption{Unit cell geometry. (a) Single layer parameterization. A JC patch is shown inside one period $P$, slab width $w$ and a variable leg length $W$. (b) Five-layer HMS unit cell formed by vertically stacking JC copper patterns, where each layer’s JC leg length $W_n$ and the collective electromagnetic interaction of the stack determines the resulting scattering parameters.}
    \label{fig:unitcell}
\end{figure}

We consider a PCB-compatible $N$-layer transmissive HMS designed for operation at nominal frequency $f_0$ (Fig.~\ref{fig:unitcell}). Similar to the configuration introduced in Part~I, each layer consists of a periodic array of Jerusalem-cross (JC) copper patterns parameterized by the tunable JC leg length $W$ (Fig.~\ref{fig:unitcell_a}), defined on dielectric laminates separating each pair of adjacent metallization layers (Fig.~\ref{fig:unitcell_b}). As formulated in Part~I, the SA \textsc{LAYERS} model provides a rapid, field-based prediction of the scattering response for such stacks while rigorously accounting for near-field (evanescent) coupling~\cite{Marcus2026LAYERS}.

To interface with the discrete generative modeling required for the inverse-design stage of our methodology (see Section~\ref{sec:results}), each layer parameter is discretized into a finite vocabulary $V$ of possible values. Specifically,
\begin{equation}
    \!\!\!W_n\!\in\!V\!=\!\{W_{\min}, W_{\min}\!+\!\Delta W, \ldots, W_{\max}\},\; n\!=\!1,\ldots,N\!\!
    \label{equ:W_n_definition}
\end{equation}
where \(W_{\min}\), \(W_{\max}\), and \(\Delta W\) define the range and resolution of the available geometric states. 
This quantization converts each continuous layer dimension into a discrete symbol drawn from the vocabulary \(V\), forming the token space processed by the AR generator. 
The choice of \(W_{\min}\), \(W_{\max}\) and \(\Delta W\) should correspond to physical realizability and modeling resolution, i.e., 
the minimal/maximal printable or etchable JC leg dimension on the PCB substrate, 
and the granularity required to capture meaningful variations in the transmission and phase without introducing redundant neighboring states.
Note that this discretization is required for the AR inverse generator, whereas the forward surrogate is trained as a continuous regression model over scalar $W$ values.

A key observation is that multilayer HMSs, in the particular JC based configuration proposed and considered herein (Fig.~\ref{fig:unitcell}) can be naturally represented as sequences: each $W_n$ denotes the design parameter of layer $n$ in the ordered stack, while the output, \(\mathbf{S}\), summarizes their joint scattering-response. In this view, let \(\mathbf{W}_{1:N} \!=\! (W_1,\ldots,W_N)\) denote the design sequence and \(\mathbf{S}\) be the sequence of the scattering S parameters. The electromagnetic \emph{forward problem} is
$\mathbf{S} =f_{\mathbf{S}}(\mathbf{W}_{1:N})$,
namely, predicting the scattering response for a given geometry.\footnote{Note that due to the efficacy and flexibility of the algorithms presented in this part, we allow ourselves to extend the available degrees of freedom and consider also asymmetric impedance sheet stacks (e.g., $W_1\neq W_{N}$). While for ideal (lossless, single-operating-frequency) HMSs, such an extension should not form additional paths to increase unit cell transmissivity for a specified transmission phase \cite{Monticone2013Full,Epstein2016TAP}, in more practical cases (non-negligible dielectric and conductor loss, operation across a band of frequencies) it may enable generation of new and useful solutions. While such symmetry breaking necessarily introduces bianisotropy \cite{Pfeiffer2014Bianisotropic, Wong2015Reflectionless, Asadchy2016Perfect, Epstein2016TAP}, this is not expected to affect negatively the performance of designs in which wave-impedance mismatch can be typically ignored (e.g., metalenses).}

More generally, the formulation above does not restrict the scattering response
$\mathbf{S}$ to a single operating condition. Instead, $\mathbf{S}$ may be viewed
as a structured sequence indexed over auxiliary dimensions such as frequency,
incidence angle, or polarization state. In this case, the forward surrogate is
tasked with learning a conditional mapping
$f_{\mathbf{S}}(\mathbf{W}_{1:N}, \mathbf{c})$, where $\mathbf{c}$ denotes a set of
conditioning variables that specify the desired operating regime. This abstraction
enables the same sequence model to be extended to richer electromagnetic objectives
without altering its core architecture, a property that is exploited in Section~\ref{subsec:broadband} to model broadband frequency responses, where $\mathbf{c}$ corresponds to discrete frequency indices spanning the operating band.

In contrast to the forward problem, the inverse task seeks one or more layer sequences 
$\mathbf{W}_{1:N}$ that produce a desired response $\mathbf{S}^\ast$. 
However, the mapping $\mathbf{S} \mapsto \mathbf{W}_{1:N}$ is generally non-bijective as 
many distinct multilayer configurations can yield similar or even identical scattering responses. 
This intrinsic one-to-many structure compounded by the complex electromagnetic interaction among layers renders direct inversion fundamentally ill-posed. To overcome this difficulty, we adopt an AR formulation inspired by language models: just as such models predict the next word from its context~\cite{Brown2020GPT3}, we seek to predict the next layer token conditioned on the previous layers and the target $\mathbf{S^\star}$.

Both the forward and inverse problems defined above involve processing ordered layer sequences whose global electromagnetic response emerges from cumulative interlayer interactions. 
As mentioned in Section~\ref{sec:introduction}, this structure naturally motivates the use of SSMs, where a latent state progressively aggregates global context as the sequence is traversed. 
In a generic discrete state space formulation, an input $\mathbf{x}_n$ updates an internal hidden state $\mathbf{h}_n$ and produces an output $\mathbf{y}_n$ according to
\begin{equation}
\mathbf{h}_n=\mathbf{A}_n\mathbf{h}_{n-1}+\mathbf{B}_n\mathbf{x}_n,\qquad
\mathbf{y}_n=\mathbf{C}_n\mathbf{h}_n,
\label{eq:ssm_generic}
\end{equation}
with learned operators $\mathbf{A}_n,\mathbf{B}_n,\mathbf{C}_n$ that vary with $n$ in the  SSM setting\cite{Gu2024Mamba}.
Crucially, the forward and inverse tasks impose different structural requirements on how sequence information should be processed. The forward problem is a deterministic regression task, in which the scattering response $\mathbf{S}$ depends on all layers through mutual electromagnetic coupling. Accordingly, forward prediction benefits from bidirectional processing that allows information to flow symmetrically across the layer index. In contrast, the inverse problem is intrinsically ill-posed and one-to-many: valid designs must be constructed sequentially, while maintaining consistency with both the previously generated layers and the target response. This naturally favors a causal, AR formulation, in which each output $\mathbf{y}_n$ corresponds to a conditional distribution over the next layer parameter. Based on these considerations, we instantiate the state space prior using Mamba, employing a bidirectional variant for forward surrogate modeling and a causal AR variant for inverse generation. 

\subsection{Detailed Pipeline}
\label{subsec:detailed_pipeline}
\begin{figure}[!t]
    \centering
    \includegraphics[width=0.48\textwidth]{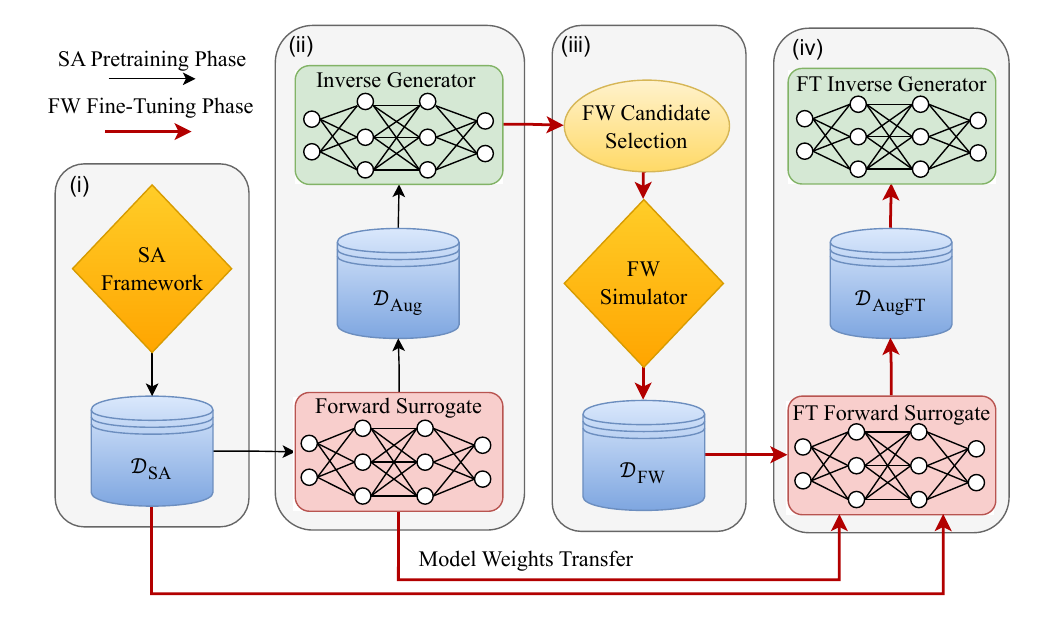}
    \caption{MetaMamba pipeline.
    (i) A large synthetic dataset $\mathcal{D}_{\mathrm{SA}}$ is generated using the 
    SA model, establishing the foundation for data-driven learning. 
    (ii) The Bi-Mamba forward surrogate is pretrained on $\mathcal{D}_{\mathrm{SA}}$, 
    used to generate an augmented dataset $\mathcal{D}_{\mathrm{Aug}}$, and an 
    initial AR-Mamba inverse generator is trained on this surrogate-produced data. 
    (iii) The pretrained inverse model generates candidate unit cell geometries; 
    high-transmission representatives are selected and evaluated with full-wave (FW) 
    simulations to form the calibration set $\mathcal{D}_{\mathrm{FW}}$. 
    (iv) The forward surrogate is fine-tuned (FT) using $\mathcal{D}_{\mathrm{FW}}$ and used to synthesize an augmented high-fidelity corpus $\mathcal{D}_{\mathrm{AugFT}}$, 
    enabling training of the final FW-calibrated AR-Mamba inverse model.}
    \label{fig:pipeline}
\end{figure}

Equipped with these observations, we proceed as laid out in Section \ref{sec:introduction}; to establish a scalable and data-efficient solution, we formulate a hybrid framework combining SA modeling, limited high-fidelity simulation, and sequence-aware generative learning (see Fig.~\ref{fig:pipeline}). For clarity, the workflow depicted in Fig.~\ref{fig:pipeline} may be viewed at two complementary levels. At the procedural level it consists of four steps (i)–(iv), whereas at the conceptual level these steps naturally fall into two training phases: a \emph{SA pretraining phase} (steps (i)–(ii), thin black arrows), in which the learning models are exposed only to SA data; and a subsequent \emph{FW fine-tuning phase} (steps (iii)–(iv), thick red arrows), in which the surrogate is calibrated to full-wave simulations while retaining the broad generalization learned from the SA model, forming the high-fidelity basis for the final inverse-model training.

As shown in step (i) of Fig.~\ref{fig:pipeline}, the hybrid framework begins by generating a large synthetic dataset $\mathcal{D}_{\mathrm{SA}}$ using the rapid 
SA model. This dataset provides the foundation for pretraining the forward 
surrogate, enabling it to capture the global scattering trends across the design space.

\begin{figure}[!t]
    \centering
    \includegraphics[width=0.48\textwidth]{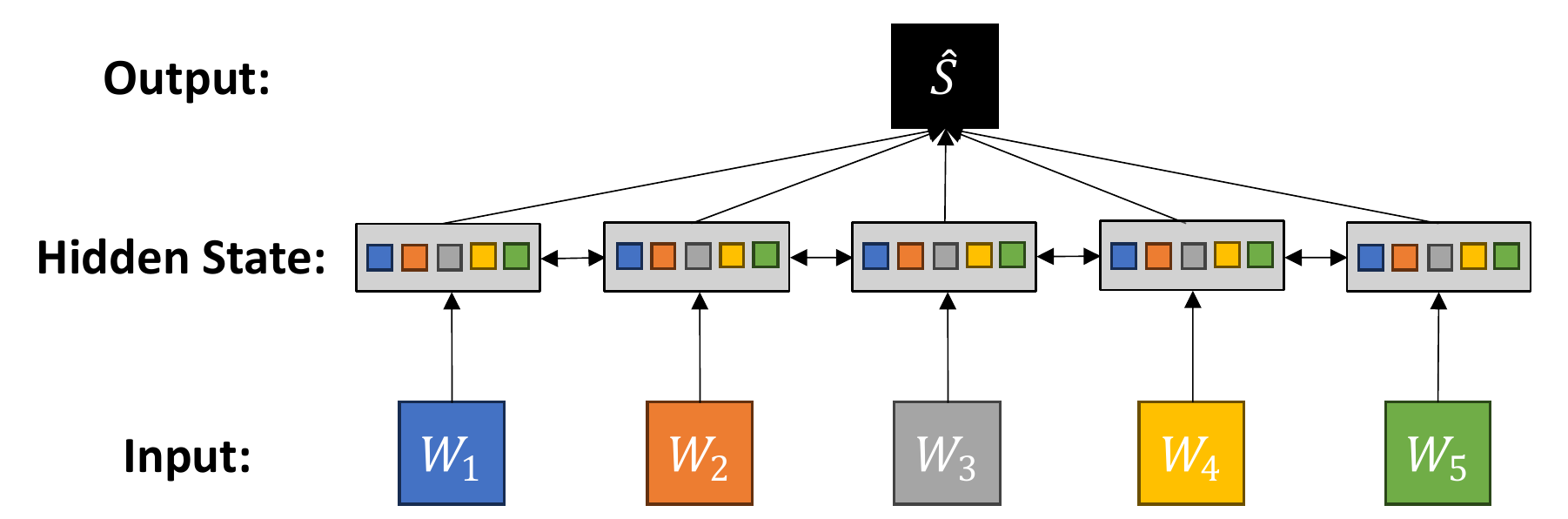}
    \caption{Bi-Mamba forward surrogate (Fig. \ref{fig:pipeline}, steps (ii), (iv), pink blocks). 
    Left-to-right and right-to-left scans are fused to capture global interlayer coupling and predict the scattering response $\hat{\mathbf{S}}$. The architecture efficiently models multilayer interactions by propagating context across all layer elements.}
    \label{fig:Bi-Mamba}
\end{figure}

The next component of the pipeline (step (ii)) starts with training a Bi-Mamba forward surrogate \(f_{\mathbf{S}} : \mathbf{W}_{1:N} \mapsto \hat{\mathbf{S}}\), trained to approximate the forward map illustrated in Fig.~\ref{fig:Bi-Mamba}. In this state space formulation, the sequence of layer parameters is processed through two recurrent directions, forward and backward, whose hidden states evolve according to learned state transitions with input-dependent modulation, allowing the model to emphasize the most informative layers while maintaining linear-time evaluation. This bidirectional flow emulates the mutual multiple-scattering interactions between layers, allowing the surrogate to propagate information across the stack with linear-time efficiency. The model is first optimized on \(\mathcal{D}_{\mathrm{SA}}\)
with standard mean-square error (MSE) loss function:
\begin{equation}
\label{eq:fwd_loss}
    \mathcal{L}_{\mathrm{fwd}}
    = \frac{1}{d_{\mathbf{S}}}
      \sum_{i=1}^{d_{\mathbf{S}}}
      \left( S_i^\star - \hat{S}_i \right)^2 .
\end{equation}
where $d_{\mathbf{S}}$ is the dimension of the $\mathbf{S}$ sequence. This procedure (Fig.~\ref{fig:pipeline}, step (ii), pink block) enables the surrogate to capture the global scattering trends learned from \textsc{LAYERS} across the design space.

\FloatBarrier

\begin{figure}[!t]
    \centering
    \includegraphics[width=0.48\textwidth]{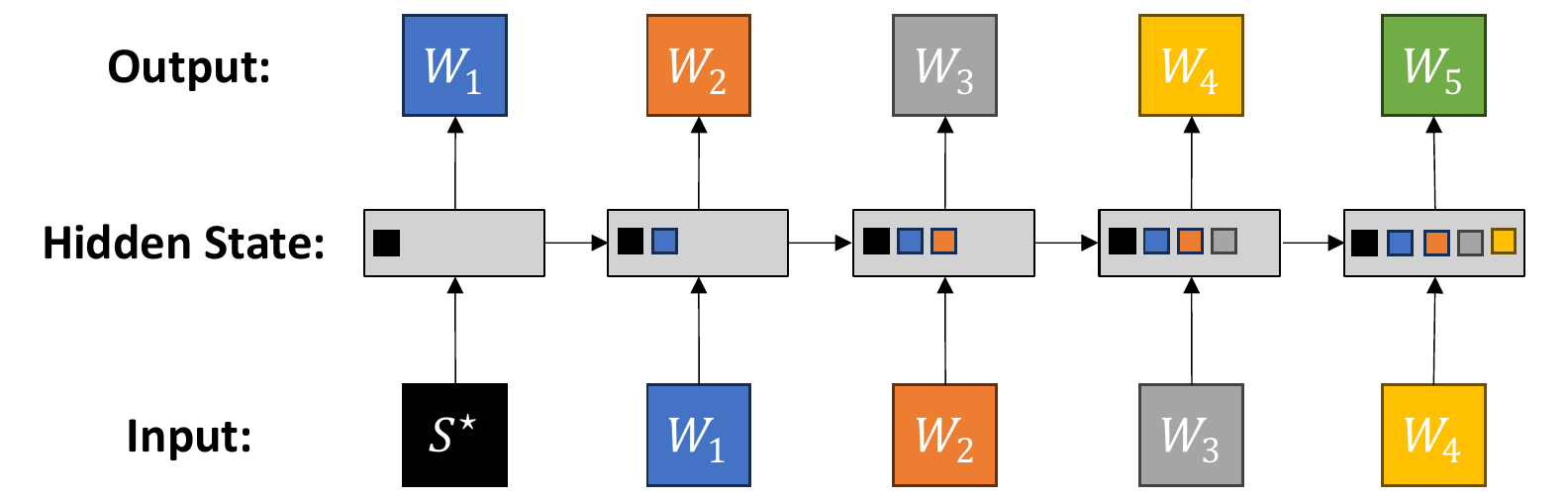}
    \caption{AR-Mamba inverse generator (Fig. \ref{fig:pipeline}, steps (ii), (iv), green blocks). 
    Conditioned on the target response $\mathbf{S^\star}$, the model predicts layer tokens 
    sequentially from left to right. At each step $n$, the next token $W_n$ is generated 
    causally from the previous tokens $\mathbf{W}_{<n}$ and the conditioning sequence $\mathbf{S^\star}$. This AR formulation resembles language models, enabling diverse sequence generation for HMS design.}
    \label{fig:ARMamba}
\end{figure}

Having the forward surrogate trained, we use it to generate a large dataset $\mathcal{D}_\mathrm{Aug}$ via Sobol sampling \cite{Sobol1967}, ensuring dense coverage of feasible scattering responses for inverse model training (Fig.~\ref{fig:pipeline}, step (ii), green block). The inverse model learns the probability function:
\begin{equation}
    p_W(\mathbf{W_{1:N}}|\mathbf{S^\star}) = \prod_{n=1}^{N} p_W(W_n|\mathbf{W_{<n}},\mathbf{S^\star}),
\end{equation}
using a causal AR-Mamba. The sequence begins with a projection of \(\mathbf{S^\star}\) into a high-dimension embedding space, followed by AR token prediction, illustrated in Fig.~\ref{fig:ARMamba}. In its training process, teacher forcing\cite{williams1989learning} is applied, opting to minimize the cross-entropy:
\begin{equation}
\mathcal{L}_{\mathrm{inv}} \;=\; -\sum_{n=1}^{N} \log p\!\left(W_n \,\middle|\, \mathbf{W}_{<n}^\star,\, \mathbf{S}^\star\right).
\end{equation}
To evaluate the inverse model performance, a reconstruction mean absolute error (MAE) metric employs the surrogate to quantify the match between the scattering response $\hat{\mathbf{S}}$ of the inverse prediction $\mathbf{\hat{W}_{1:N}}$ and the desired scattering $\mathbf{S^\star}$:
\begin{equation}
\label{eq:reconstruction_loss}
    \mathcal{L}_{\mathrm{rec}}
    = \frac{1}{d_{\mathbf{S}}}\sum_{i=1}^{d_{\mathbf{S}}}
    \big| S^\star_i - \hat{S}_i \big|,
\end{equation}
Since not any desired \(\mathbf{S^\star}\) is physically realizable, the inverse model aims to approximate the feasible distribution and return designs close to \(\mathbf{S^\star}\). If \(\mathbf{S^\star}\) lies far outside this distribution, the resulting sequence may be unrealizable, which can be readily identified by evaluating it through the surrogate model.

Designs are generated by the inverse model using either top-$k$ or top-$p$ sampling~\cite{Fan2018,Holtzman2019}, then re-ranked by the calibrated surrogate $f_{\mathbf{S},\mathrm{cal}}$ to form a LUT of unit cells geometries $\mathbf{W}_{1:N}$ that realize the desired response \(\mathbf{S^\star}\). In \emph{top-$k$ sampling}, the next token is drawn randomly from the $k$ most likely candidates, while in \emph{top-$p$ sampling}, it is drawn from the smallest set of candidates whose cumulative probability exceeds $p$. This concludes step~(ii) of Fig.~\ref{fig:pipeline}, resulting in an initial inverse generator trained solely on the surrogate-augmented dataset.

\begin{figure}[!t]
    \centering
    \includegraphics[width=0.48\textwidth]{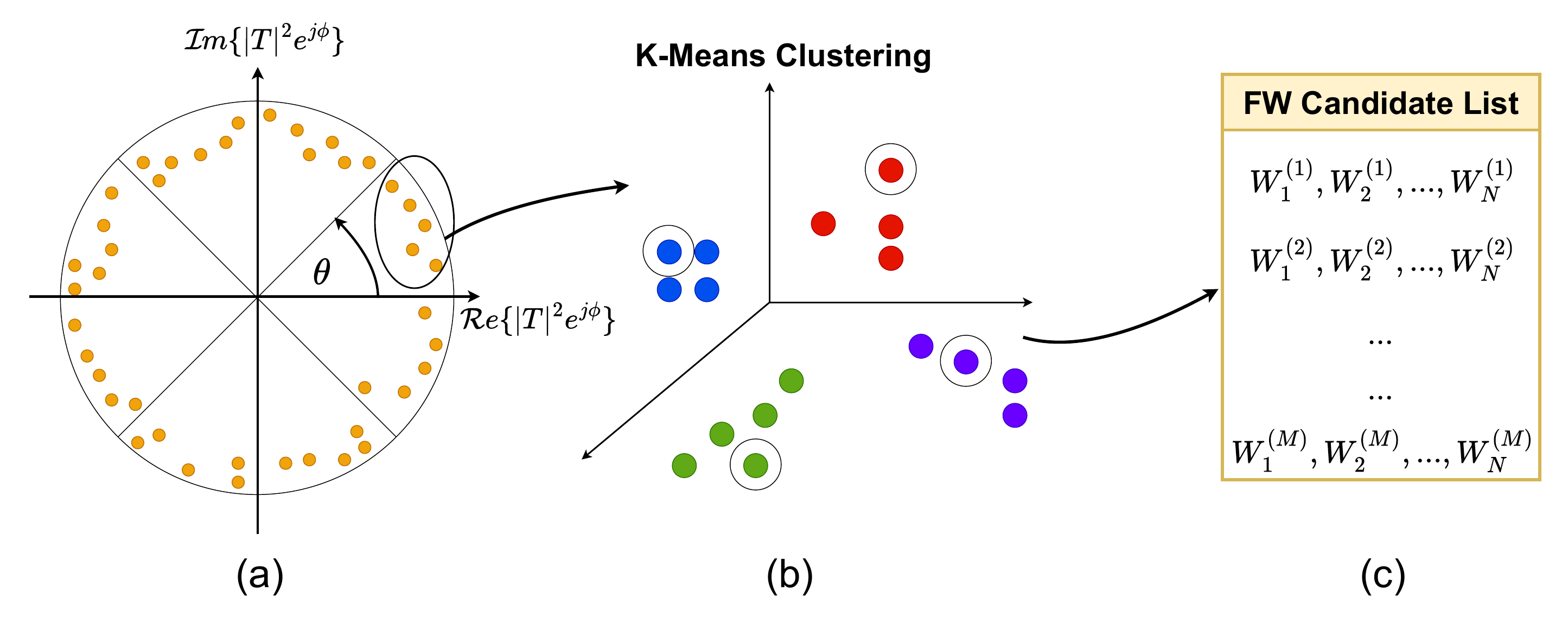}
    \caption{Candidate selection strategy. 
    (a) The pretrained inverse model generates a large pool of high-transmission 
    candidates, whose predicted responses $(|T|^2,\phi)$ are visualized in 
    polar form and partitioned into fixed phase sectors. 
    (b) Within each phase sector, candidates are clustered in the geometric parameter 
    space $\mathbf{W}_{1:N}$ using K-means to promote diversity.
    (c) The highest-transmission representative from each cluster (encircled in (b)) is selected to form the compact, diverse calibration set that together with the full-wave simulation results will form $\mathcal{D}_{\mathrm{FW}}$.}
\label{fig:candidate_selection}
\end{figure}

To bridge to high-fidelity physics and realize step~(iii), $M$ candidate unit cells ($M$ specific combinations of $\mathbf{W}_{1:N}$ for the configuration in Fig. \ref{fig:unitcell}) are selected to be simulated in general purpose electromagnetic solver (CST). The candidate selection process, illustrated in Fig.~\ref{fig:candidate_selection}, follows a general two-stage strategy. First, the pretrained inverse model (Fig.~\ref{fig:pipeline}, step (ii), green block), conditioned on properties of interest (in our case, high power transmission efficiency combined with broad phase coverage), generates a large set of unit cell geometries. Each candidate is then evaluated by the surrogate model to obtain its predicted scattering properties (e.g $(|T|^2,\phi)$), which we visualize in polar form to reveal how generated designs populate the magnitude–phase space (Fig. \ref{fig:candidate_selection}a). To ensure that the full-wave calibration set spans the entire target range while avoiding redundant geometries, we partition the candidates into fixed phase sectors of size $\theta$ and apply K-means clustering \cite{MacQueen1967} in the geometric parameter space $(W_1,\ldots,W_N)$ within each sector (Fig. \ref{fig:candidate_selection}b). Although the choice of phase sectors is oriented to the present task, the same approach applies more broadly by binning candidates according to any task-relevant response dimension before clustering in geometry space. Following that, from every cluster we select the highest-transmission representative to form a compact, geometrically diverse set of full-wave simulations denoted $\mathcal{D}_{\mathrm{FW}}$ (Fig. \ref{fig:candidate_selection}c).

\emph{Full-wave calibration is the key step that unlocks the hybrid scheme’s strength in the accuracy–efficiency trade-off.} Rather than training from scratch on scarce full-wave data, in step (iv) of Fig.~\ref{fig:pipeline} we continue the optimization with the forward surrogate $f_{\mathbf{S}}$~\cite{yosinski2014transferable} that was pretrained on \(\mathcal{D}_{\mathrm{SA}}\), and interleave SA and full-wave batches during fine-tuning (a rehearsal strategy that mitigates catastrophic forgetting~\cite{french1999catastrophic}, corresponding to the process denoted in red thick arrows feeding the pink block of step (iv) in Fig.~\ref{fig:pipeline}). The calibration objective mixes the two fidelity levels:
\begin{equation}
\mathcal{L}_{\mathrm{cal}} \;=\; \lambda_{\mathrm{FW}}\,\mathbb{E}_{\mathcal{D}_{\mathrm{FW}}}\!\big[\mathcal{L}_{\mathrm{fwd}}\big]
\;+\; \lambda_{\mathrm{SA}}\,\mathbb{E}_{\mathcal{D}_{\mathrm{SA}}}\!\big[\mathcal{L}_{\mathrm{fwd}}\big],
\label{eq:calib}
\end{equation}
where \(\lambda_{\mathrm{FW}}\) and \(\lambda_{\mathrm{SA}}\) control the balance between high-accuracy anchoring and broad generalization. This preserves the global coverage learned from \(\mathcal{D}_{\mathrm{SA}}\) while aligning the predictions with high-fidelity “ground truth.”

After calibration, the surrogate \({f}_\mathbf{S,cal}\) synthesizes an augmented corpus 
\(\mathcal{D}_{\mathrm{AugFT}}\!=\!\{(\mathbf{W}_{1:N}, f_\mathbf{S,cal}(\mathbf{W}_{1:N}))\}\) with near full-wave accuracy, enabling efficient training of a calibrated, accurate AR-Mamba inverse model (Fig.~\ref{fig:pipeline}, step (iv), green block). At this stage, the MetaMamba pipeline is complete, with the calibrated surrogate enabling a final high-fidelity inverse generator.

\section{Results and Discussion}\label{sec:results}
\label{sec:results}

\subsection{HMS Configuration and SA-Model Accuracy}
\label{subsec:HMSconfig}

We apply the MetaMamba framework developed in Section~\ref{sec:method} to a representative multilayer transmissive HMS configuration operating at 20~GHz. The ultimate goal is to construct a high-fidelity, fabrication-ready LUT of five-layer dual-polarized unit cells, covering the full \(2\pi\) transmission phase range with maximum possible efficiency. 

As noted in Section~\ref{sec:method}, the physical configuration under study, shown schematically in Fig.~\ref{fig:unitcell}, comprises five stacked metallic layers featuring JC patterns, each defined by a horizontal copper trace of variable leg length \(W_n\) (the $n$th JC leg-length), supported by dielectric substrates and bond layers. The geometry and material parameters of the meta-atoms as used in the SA model are summarized in Table~\ref{tab:geometry_constants}~\cite{Marcus2026LAYERS}. Correspondingly,  
the minimum and maximum allowable JC leg lengths are \(W_\mathrm{min} = 0\)~mil and \(W_\mathrm{max} = 80\)~mil, respectively, sampled with \(\Delta W = 1\)~mil resolution, essentially composing a vocabulary \(V\) of 81 discrete values and a design space of size \(|V|^5 = 81^5 \approx 3.5\times10^9\) possible layer combinations. The choice of $\Delta W = 1$ mil reflects standard PCB fabrication capabilities and ensures that neighboring tokens correspond to smoothly varying scattering responses.

\begin{table}[t]
\renewcommand{\arraystretch}{1.2}
\caption{Geometry and Material Constants Used in SA Model~\cite{Marcus2026LAYERS}}
\label{tab:geometry_constants}
\centering
\begin{tabular}{@{}lcl@{}}
\toprule
\textbf{Parameter} & \textbf{Value} & \textbf{Material / Property} \\
\midrule
Period, $d$ & 3.06~\text{mm} & $\approx \lambda_0/4.9$ at 20~GHz \\
Wavelength, $\lambda_0$ & 14.99 mm & At 20~GHz \\
Cross width, $w$ & 4 mil & Copper \\
Distance from cell edge $w_s/2$ & 2 mil & --- \\
Copper thickness & $18~\mu\text{m}$ & $\sigma = 5.8 \times 10^7$ S/m \\
Substrate thickness & 30 mil & \makecell[l]{Isola Astra MT77,\\ $\varepsilon_r=3$, $\tan\delta=0.001$} \\
Bond layer thickness & 2 mil & \makecell[l]{Isola Astra MT77,\\ $\varepsilon_r=3$, $\tan\delta=0.001$} \\
\bottomrule
\end{tabular}
\end{table}

\begin{figure}[t]
    \centering    
    \subfloat[]{%
        \includegraphics[width=0.48\linewidth]{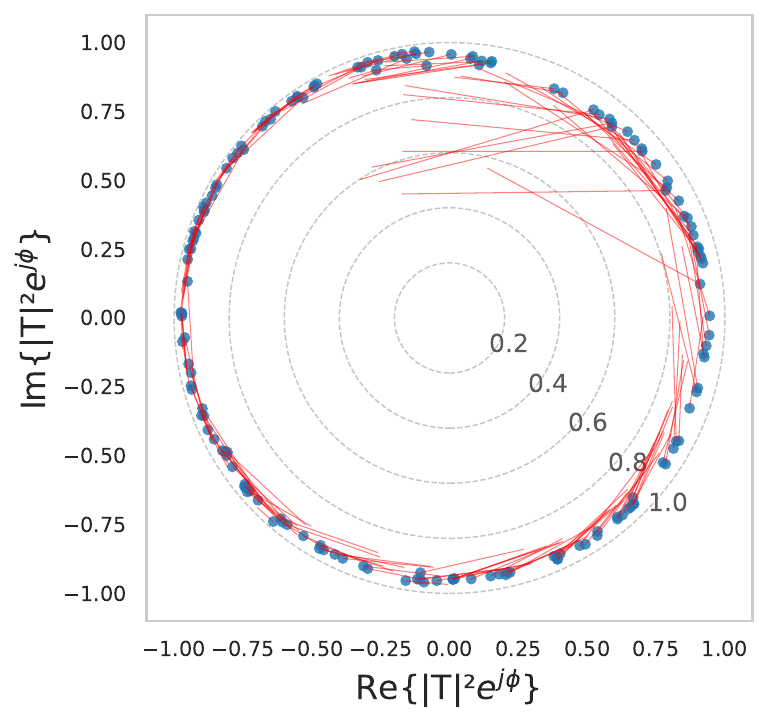}%
        \label{fig:polar_error_layers}
    }
    \hfill 
    \subfloat[]{%
        \includegraphics[width=0.48\linewidth]{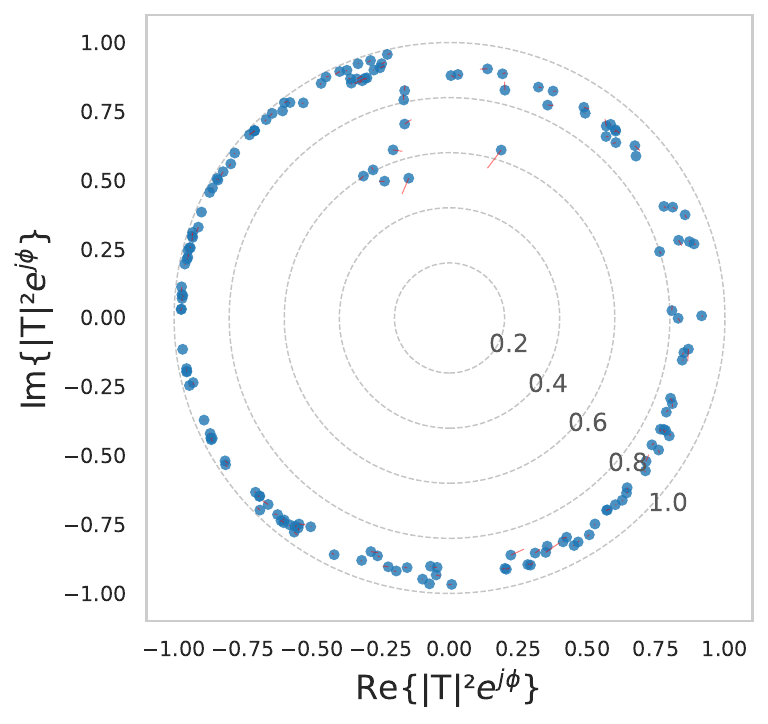}%
        \label{fig:polar_error_forward}
    }
    \caption{Polar error maps comparing CST ground truth with (a) the SA LAYERS model and (b) the calibrated forward surrogate. Each blue marker denotes a predicted complex transmittance $|T|^2e^{j\phi}$ for a held-out test configuration used in the calibration study (Sec.~\ref{subsec:forwardcalib}). Red line segments connect each prediction to its corresponding CST result in the complex plane; their length therefore directly indicates the magnitude of the prediction error. Panel (a) reveals pronounced, phase-dependent structural discrepancies in the LAYERS model, while panel (b) shows that these deviations are largely eliminated after calibration.}
    \label{fig:polar_error_maps}
\end{figure}

Before proceeding to the MetaMamba learning pipeline (Fig.~\ref{fig:pipeline}, steps (ii)–(iv)), it is instructive to first assess the intrinsic accuracy of the underlying SA LAYERS model on which the forward surrogate is initially pretrained (step (i)). To this end, Fig.~\ref{fig:polar_error_layers} presents a polar error visualization comparing LAYERS predictions against CST ground truth for a held-out set of unit cell configurations.
Importantly, the points shown in Fig.~\ref{fig:polar_error_maps} do \emph{not} constitute a lookup table (LUT), nor are they the result of an exhaustive design sweep. Rather, they correspond to the held-out test set used in the forward-model fine-tuning procedure described in Sec.~\ref{subsec:forwardcalib}, and are shown here as a diagnostic preview to highlight the necessity of calibration. Each blue marker on the unit circle represents the complex transmission coefficient ($|T|^2 e^{j\phi}$) predicted either by the LAYERS model (a) or by the calibrated forward surrogate (b). For each configuration, a red line connects the prediction to its corresponding CST result in the complex plane; the length of this line therefore directly reflects the prediction error.
As seen in Fig.~\ref{fig:polar_error_layers}, the LAYERS predictions exhibit systematic, phase-dependent deviations from CST that are clearly structural rather than random. In several regions of the polar plane, these discrepancies manifest as large angular offsets, sometimes accompanied by noticeable degradation in $|T|^2$. Such discrepancies would directly mislead any inverse design model trained solely on LAYERS-generated data,motivating the calibration stage introduced next in Section.~\ref{subsec:forwardcalib} to bring the forward-model predictions into alignment with CST ground truth.

\subsection{Forward Surrogate Calibration}
\label{subsec:forwardcalib}

To implement and subsequently evaluate the performance of the forward surrogate, we follow the two-phase training pipeline described in Section~\ref{sec:method} (Fig.~\ref{fig:pipeline}). The Bi-Mamba~\footnote{Full architectural details of the Bi-Mamba forward model, including layer dimensions and parameter counts, are provided in Appendix~\ref{app:architectures} (Table~\ref{tab:forward_model_details}). The corresponding training schedule and hyperparameters used in this calibration phase are summarized in Appendix~\ref{app:hyperparams_config} (Table~\ref{tab:training_hparams}).} model is first pretrained on a dataset of $\approx{524,000}$ SA generated examples ($\mathcal{D}_{\mathrm{SA}}$) to learn broad scattering trends, enabling efficient modeling of the layer-to-response mapping (steps (i) and (ii) in Fig. \ref{fig:pipeline}). Next, to mitigate the residual inaccuracies of the SA model exemplified in Fig.~\ref{fig:polar_error_layers}, we refine the surrogate by fine-tuning it on a limited number of full-wave simulations conducted in CST Microwave Studio under periodic Floquet boundaries (steps (iii) and (iv)).

To select informative and representative samples for this calibration step (step (iii)), we follow the steps depicted in Fig.~\ref{fig:candidate_selection} and generate $1080$ candidate unit cells using the AR inverse generator (Fig.~\ref{fig:pipeline}, step (ii), green block) described in Section~\ref{sec:method}. The chosen sampling strategy was top-$k$, with $k=20$. These cell geometries are mapped to their corresponding scattering responses using the forward surrogate that was trained on SA data (Fig.~\ref{fig:pipeline}, step (ii), pink block). We then divide the results into $\theta=5^\circ$ sectors, and apply K-means clustering (with $K=15$) over the design geometry sequence \((W_1,\dots,W_5)\). The geometries with the highest transmission in each cluster are selected, and their full-wave responses are simulated using CST, a total of $1080$ simulations that form the dataset $\mathcal{D}_{\mathrm{FW}}$ (Fig.~\ref{fig:pipeline}, step (iii)). This clustering-based sampling ensures that the calibration set spans a diverse and meaningful range of electromagnetic behaviors, maximizing fine-tuning impact while minimizing simulation budget. Finally, The calibrated model $f_{\mathbf{S},\mathrm{cal}}$ (pink block in step (iv)) is optimized using the loss $\mathcal{L}_{\mathrm{cal}}$ of \eqref{eq:calib}; the complete calibration schedule and associated hyperparameters are listed in Appendix~\ref{app:ft_hparams}.

In Table~\ref{tab:fwd_mse} we report the forward loss $\mathcal{L}_\mathrm{fwd}$ of~\eqref{eq:fwd_loss}, computed as the mean squared error (MSE) over the normalized targets $[\sin\phi,\cos\phi,|T|^2]$, evaluated on both the SA and CST test sets. The phase is encoded through its sine and cosine components to avoid wrap-around discontinuities and to ensure a smooth, continuous representation, followed by normalization to the range [0,1] in order to promote stable and balanced training.
Several important trends emerge directly from the table. First, although the SA-trained forward model achieves low error on the SA test set ($2.3\times10^{-4}$), i.e., compared to the results produced by LAYERS for the examined $\textbf{W}$ case studies, its error increases by nearly two orders of magnitude when evaluated against CST ground truth ($1.9\times10^{-2}$), reflecting the inherent modeling inaccuracies of the SA solver. After fine-tuning, the calibrated surrogate deliberately sacrifices some agreement with the SA data—its SA test error increases modestly to $4.5\times10^{-4}$—yet simultaneously achieves a dramatic improvement on CST, reducing the CST test error by a factor of $260$ to $7.3\times10^{-5}$. This behavior indicates that fine-tuning (Fig.~\ref{fig:pipeline}, step (iv)) successfully shifts the surrogate toward the true full-wave physics and confirms that catastrophic forgetting did not occur.
These numerical trends corroborate the qualitative behavior observed in Fig.~\ref{fig:polar_error_maps}. Noticeably, the shrinkage of the red lines in Fig.~\ref{fig:polar_error_forward}, visibly shows the suppression of the systematic phase and magnitude deviations present in the raw SA predictions in Fig.\ref{fig:polar_error_layers}. Together, the results confirm that full-wave calibration is indispensable for aligning the forward surrogate with CST-level accuracy, enabling reliable, high-fidelity inverse design.

\begin{table}[!t]
\centering
\caption{Test MSE of forward surrogates on low-fidelity (LF) SA and high-fidelity (HF) CST datasets.}
\renewcommand{\arraystretch}{1.15}
\begin{tabular}{lcc}
\hline
 & SA Test & CST Test \\
\hline
Forward Surrogate & $2.3 \times 10^{-4}$ & $1.9 \times 10^{-2}$ \\
FT Forward Surrogate & $4.5 \times 10^{-4}$ & $7.3 \times 10^{-5}$ \\
\hline
\end{tabular}

\vspace{2mm}
\raggedright
Forward Surrogate corresponds to Fig.~\ref{fig:pipeline}, step (ii);  
Fine-tuned (FT) Forward Surrogate corresponds to Fig.~\ref{fig:pipeline}, step (iv).
\label{tab:fwd_mse}
\end{table}

As a sanity check, we also trained a forward surrogate using only the CST calibration data, without SA pretraining. This model achieved a CST test MSE of $9\times10^{-4}$, which may appear competitive in isolation, yet remained inferior to the SA-pretrained and CST-calibrated surrogate. When evaluated on the global LAYERS test set, however, the CST-only model exhibited a substantially higher error of $1.3\times10^{-2}$, indicating that its apparent accuracy is confined to a narrow region of the design space. This behavior is expected given the limited diversity and strong structural bias of the CST calibration set, which predominantly contains high-efficiency designs; training on this data alone yields a model that interpolates locally but lacks global geometric–response context. By contrast, while the SA-only surrogate incurs higher error on CST—particularly in regions where it is least confident—it captures broad trends across the full design space. The calibrated surrogate combines both advantages: SA pretraining provides global coverage and physical context, while CST calibration corrects systematic inaccuracies in the high-efficiency regime, resulting in uniformly superior performance across all evaluation metrics.

\begin{figure}[t]
  \centering
  \subfloat[]{%
    \includegraphics[width=0.48\linewidth]{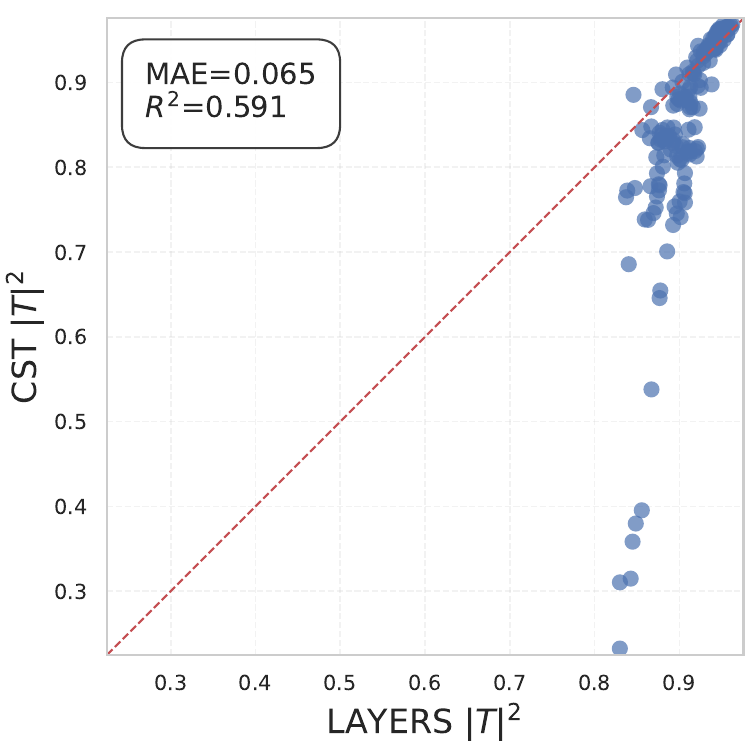}
  }\hfill
  \subfloat[]{%
    \includegraphics[width=0.49\linewidth]{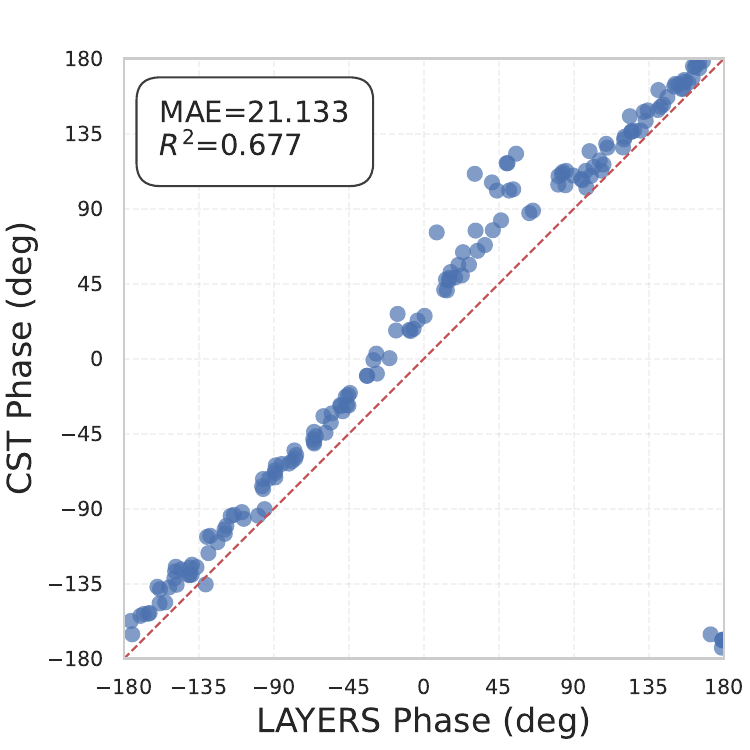}
  }\\[-4mm]
  \subfloat[]{%
    \includegraphics[width=0.48\linewidth]{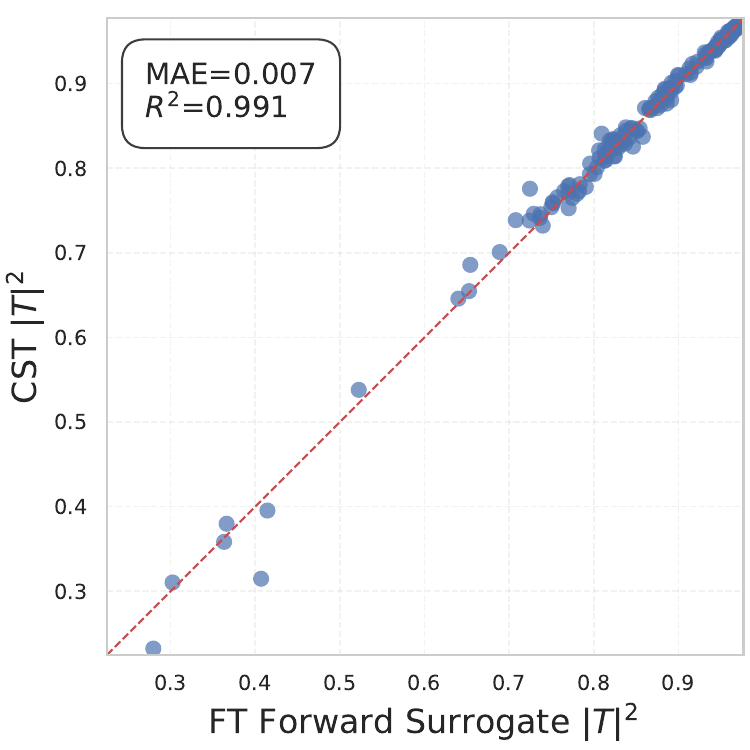}
  }\hfill
  \subfloat[]{%
    \includegraphics[width=0.49\linewidth]{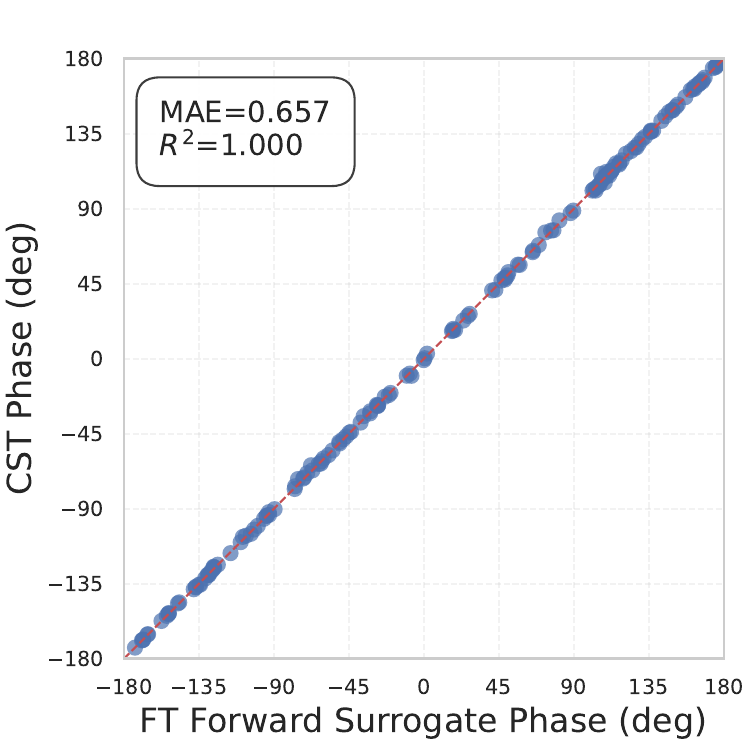}
  }\\[-2mm]
  \caption{LAYERS and the fine-tuned forward surrogate vs. CST on the held-out test set. Parity plots of transmittance~$|T|^2$ and phase~$\phi$ compare SA-based predictions and the calibrated forward surrogate with CST results. Panels (a,b) correspond to LAYERS vs. CST, while (c,d) show the calibrated surrogate vs. CST. Post-calibration agreement improves substantially, with $\mathcal{R}^2$ values increasing from $0.59$ to $0.99$ for $|T|^2$ and from $0.67$ to $1.0$ for phase.}
  \label{fig:FWD_SA_CST_parity_plots}
\end{figure}

These observations are further reinforced by the data presented in Table~\ref{tab:fwd_cst_agreement}, comparing the SA model (LAYERS) and the calibrated forward surrogate (Fig~\ref{fig:pipeline}, step (iv), pink block) against a held out test set from the CST simulation dataset. The calibrated forward surrogate reduces efficiency error by an order of magnitude and phase error by more than $30\times$, demonstrating the benefit of the calibration process (steps (iii) and (iv)).
This contrast is also evident in the parity plots of Fig.~\ref{fig:FWD_SA_CST_parity_plots}. Panels (a,b), corresponding to LAYERS predictions versus CST, exhibit substantial scatter and systematic deviations in both transmittance magnitude and phase, reflected in moderate $\mathcal{R}^2$ values of $0.591$ and $0.677$ respectively. In contrast, panels (c,d), which show the calibrated forward surrogate versus CST, display points tightly clustered along the diagonal, with post-calibration $\mathcal{R}^2$ scores approaching unity in both magnitude and phase. Here, $\mathcal{R}^2$ denotes the coefficient of determination, computed as the squared Pearson correlation between LAYERS or surrogate predictions and CST references; values close to unity indicate that both efficiency and phase variations are accurately captured across the design space, establishing the superiority of the calibrated model over the SA model in this regard.

\begin{table}[t]
\centering
\caption{Agreement with CST on the test set.}

\renewcommand{\arraystretch}{1.1}
\begin{tabular}{lcc}
\hline
 & Efficiency error & Phase$^a$ error \\
\hline
LAYERS$^b$ (mean) & 0.065 (0.067) & 21.13$^\circ$ (21.07$^\circ$) \\
FT Forward Surrogate (mean) & 0.0067        & 0.656$^\circ$ \\
\hline
LAYERS$^b$ (std)  & 0.102 (0.092) & 13.23$^\circ$ (12.97$^\circ$) \\
FT Forward Surrogate (std)  & 0.01         & 0.72$^\circ$ \\
\hline
LAYERS$^b$ (min)  & $\approx 0$ ($\approx 0$) & 4.5$^\circ$ (1.06$^\circ$) \\
FT Forward Surrogate (min)  & $\approx 0$               & $\approx 0$ \\
\hline
LAYERS$^b$ (max)  & 0.59 (0.65)   & 80.39$^\circ$ (80.78$^\circ$) \\
FT Forward Surrogate (max)  & 0.09          & 4.68$^\circ$ \\
\hline
\end{tabular}
\vspace{1mm}
\\
\raggedright
$^a$ Phase errors are wrapped absolute differences in degrees.\\
$^b$ For LAYERS, values outside parentheses are test-set errors, while values in parentheses are computed w.r.t.\ the full CST dataset. Efficiency errors are absolute differences in $|T|^2$.
\label{tab:fwd_cst_agreement}
\end{table}

To quantify the calibration efficiency of the Bi-Mamba forward surrogate, we compared its agreement with CST across different calibration budgets. All variants were trained from randomly selected subsets of a common pool of 1080 CST simulations and evaluated on the same strictly held-out test set of 162 candidates (15\% of the total). For each budget (270, 540, and 1080 samples), a consistent $70\%/15\%$ training/validation split was applied, ensuring direct and fair comparability across calibration sizes.

\begin{table}[t]
\centering
\caption{Calibration ablation of the forward surrogate (Cal) under varying CST budgets. Errors are absolute differences in $|T|^2$ and wrapped phase errors in degrees, evaluated on the same 162-candidate test set.}
\renewcommand{\arraystretch}{1.15}
\begin{tabular}{lcccccc}
\toprule
\textbf{Budget} & \textbf{Metric} & \textbf{Mean} & \textbf{Std} & \textbf{Min} & \textbf{Max} \\
\midrule
270 & Phase (deg) & 1.1 & 2.02 & $\approx 0$ & 19.81 \\
    & Magnitude  & 0.0125 & 0.0175 & $\approx 0$ & 0.137 \\
\midrule
540 & Phase (deg) & 0.87 & 1.685 & $\approx 0$ & 15.91 \\
    & Magnitude  & 0.0088 & 0.015 & $\approx 0$ & 0.154 \\
\midrule
1080 & Phase (deg) & 0.656 & 0.72 & $\approx 0$ & 4.68 \\
     & Magnitude  & 0.0067 & 0.01 & $\approx 0$ & 0.09 \\
\bottomrule
\end{tabular}
\label{tab:calib_ablations}
\end{table}

As summarized in Table~\ref{tab:calib_ablations}, even with only $270$ CST calibration samples, the Bi-Mamba forward surrogate achieves a mean wrapped phase error of $1.1^{\circ}$ and a mean magnitude error of $0.0125$, demonstrating strong agreement with full-wave simulations despite the limited calibration budget. Increasing the number of CST samples to $540$ and $1080$ leads to a steady and monotonic reduction in both phase and magnitude errors; however with progressively smaller accuracy improvements. While mean and standard-deviation errors decrease smoothly with increasing calibration budget, the behavior of the maximum error merits brief discussion. At smaller budgets (270 and 540 samples), the maximum phase and magnitude errors are dominated by a small number of outlier configurations that occupy sparsely populated regions of the design space. Such regions are statistically underrepresented in random calibration subsets and therefore benefit less from limited fine-tuning. As the calibration budget increases, these rare configurations are progressively incorporated, leading to a sharp reduction in worst-case errors. Importantly, even in the most challenging cases, the calibrated surrogate substantially outperforms the SA LAYERS model across all budgets (\textit{cf.} Table~\ref{tab:fwd_cst_agreement}), indicating that the remaining large errors are both localized and non-systematic. These results demonstrate that reliable, near–CST-level accuracy can already be attained with a few hundred strategically selected simulations, underscoring the data efficiency of the proposed SA-pretrained and selectively fine-tuned surrogate modeling approach.

\subsection{Inverse Design Performance}
\label{subsec:inverse_design}

Having established a high-fidelity forward surrogate (Fig.~\ref{fig:pipeline}, step (iv), pink block), we next evaluate the performance of the inverse generator \footnote{The architecture of the AR-Mamba inverse model is detailed in Appendix~\ref{app:architectures} (Table~\ref{tab:inverse_model_details}), while its training setup is described in Appendix~\ref{app:hyperparams_config}.} (step (iv), green block). The goal of the inverse model is to synthesize multilayer HMS unit cells whose scattering response matches a given target $\mathbf{S}^\star$, ideally achieving high transmission and covering the entire $2\pi$ phase range. The AR-Mamba generator was trained on a Sobol sequence of ${\approx2\times10^6}$ samples created by the calibrated forward surrogate, learning to predict layer-wise geometries conditioned on the desired response and prior layers in the stack.

During inference, the generator autoregressively predicts each layer token conditioned on the target and previously generated layers. The simplest strategy, \emph{greedy decoding}, always selects the single most likely token at each step (top-$k=1$). This yields a single deterministic design sequence, effectively providing the model’s “best guess” for a given target. Under greedy decoding, the inverse model achieves a reconstruction loss of 
$\mathcal{L}_\text{rec} = 3.9\times10^{-3}$ on the test set (\textit{cf.}~\eqref{eq:reconstruction_loss}), indicating that the model can recover CST-consistent solutions in a one-to-one fashion.

\begin{figure}[t]
  \centering
  \includegraphics[width=\linewidth]{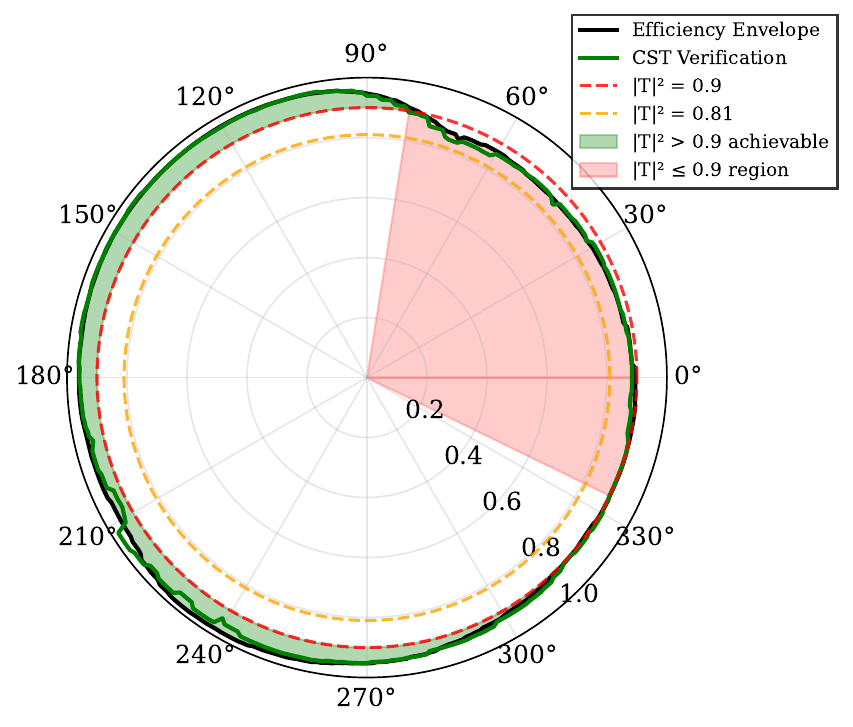}
    \caption{Feasibility envelope estimated with the calibrated inverse model. 
    The red dashed circle indicates $|T|^2 = 0.9$. 
    The envelope characterizes the maximum attainable power transmission efficiency at each phase, revealing physically realizable and non-realizable regions of the $(|T|^2,\phi)$ domain. Additional CST simulations confirm the validity of the predicted frontier.}
  \label{fig:frontier}
\end{figure}

Beyond greedy decoding, MetaMamba can exploit stochastic decoding strategies to generate multiple candidate designs. With either sampling strategy, the model can produce thousands of candidates in parallel on GPU ($\sim$8 batch decodings/sec, each batch can be in the order of $10^3$). We therefore turn to sampling-based decoding strategies and evaluate the physical boundaries of our five-layer JC design configuration. To this end, we densely probe the $(|T|^2,\phi)$ plane by sweeping $|T|^2\in[0.81,1.0]$ in $0.01$ steps and $\phi\in[0^\circ,360^\circ)$ in $2^\circ$ steps. For each target on this grid, the AR-Mamba generator produces a batch of $B=512$ candidate designs using top-$k=20$ sampling. The best-achieved efficiency per phase bin defines the \emph{feasibility envelope}—the upper boundary of achievable high-efficiency responses with the considered structure. Generation of the full envelope requires less than 10 minutes on a single GPU. The resulting frontier, shown in Fig.~\ref{fig:frontier}, reveals a nonuniform physical landscape: while many phases admit efficiencies well above $90\%$ (light green background), certain regions exhibit inherent limitations due to the restricted geometric degrees of freedom and inevitable conductor loss in realistic metals (light red background). Notably, within the estimated feasibility envelope, the minimum achievable power transmission efficiency is $|T|^2 = 0.8385$ (corresponding to $|T| = 0.916$) occurring at $\phi = 69^\circ$. This value represents the most restrictive point of the envelope and serves as a lower bound on physically realizable high-efficiency responses for the considered five-layer JC configuration. Selected points along the envelope were validated through additional CST simulations (solid green curve), confirming that the envelope accurately reflects the underlying achievable response set. Targets lying outside the envelope are considered out-of-distribution (OOD) for the current design space. For such targets, the inverse model naturally fails to produce successful designs, and either generates the nearest feasible envelope point, or we abstain if no point passes thresholds.

With the feasible response set established, we next analyze the generative model’s behavior \emph{within} this envelope. A generated design is deemed \emph{successful} if it satisfies both a phase accuracy constraint,
$|\phi_{\mathrm{pred}} - \phi^\star| \le 5^\circ$, 
and an efficiency constraint,
$|T_{\mathrm{pred}}|^2 \ge 0.95\,|T^\star|^2$, 
where $(|T^\star|^2,\phi^\star)$ denotes the desired target response and $(|T_{\mathrm{pred}}|^2,\phi_{\mathrm{pred}})$ is the response associated with the design sequence predicted by the inverse model. 

To quantify performance across stochastic decoding runs, we define several complementary metrics. First, reliability is assessed via the \emph{success rate} (SR of \eqref{eq:success_rate}), defined as the ratio between the number of successful samples, denoted by $n_s$, and the total number of generated samples in the batch $B$. Formally, the number of successful designs is given by
\begin{equation}
\label{eq:success_number}
n_s
= \sum_{i=1}^{B}
\mathbb{I}\!\left(
|\phi_{\mathrm{pred}}^{(i)} - \phi^\star| \le 5^\circ
\;\wedge\;
|T_{\mathrm{pred}}^{(i)}|^2 \ge 0.95\,|T^\star|^2
\right),
\end{equation}
where $\mathbb{I}(\cdot)$ denotes the indicator function. The corresponding success rate is then
\begin{equation}
\label{eq:success_rate}
\mathrm{SR} = \frac{n_s}{B}.
\end{equation}
Second, to capture structural multiplicity, we report the number of \emph{unique} successful design sequences, defined as the cardinality of the set of distinct geometry sequences $\{(W_1,\dots,W_5)\}$ among the $n_s$ designs satisfying the success criteria.
Finally, to quantify how broadly the inverse model explores the discrete design space, we introduce a \emph{diversity} metric, defined as the mean pairwise token $\ell_1$ distance across all successful designs:
\begin{equation}
\label{eq:diversity_eq}
\!\!\!\text{Diversity}
\!=\!\frac{2}{n_s(n_s-1)}
\!\!\sum_{i=1}^{n_s-1}\!\! \sum_{j=i+1}^{n_s}\!\!
\frac{1}{N}
\sum_{n=1}^{N}
\left| W_n^{(i)} - W_n^{(j)} \right|,
\end{equation}
where $n_s$ is of \eqref{eq:success_number} and $N$ is the layer sequence length. 

\begin{figure}[t]
  \centering
  \includegraphics[width=0.48\textwidth]{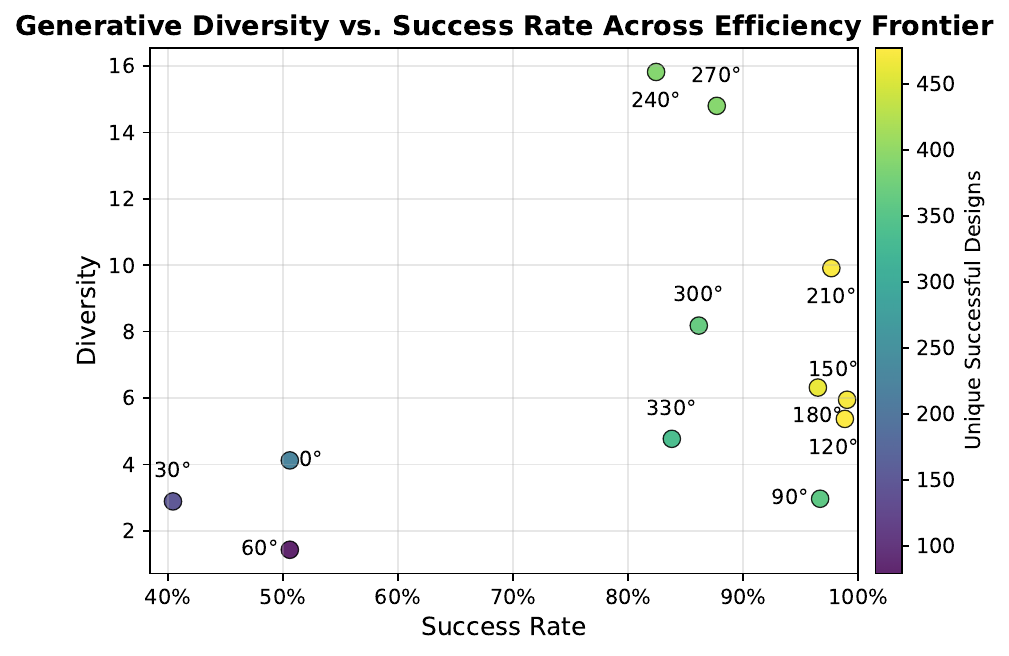}
  \caption{Diversity \eqref{eq:diversity_eq} vs. Success rate of the generated designs with top-$k=20$ decoding policy and a budget of $B=512$ samples. Each marker corresponds to a target phase across the feasibility envelope, with color encoding the number of unique successful sequences.}
  \label{fig:diversity_main}
\end{figure}

Having defined the success rate, uniqueness and diversity metrics, we now examine how these quantities vary across the feasible response space. Figure~\ref{fig:diversity_main} shows the relationship between success rate and structural diversity for top-$k=20$ decoding of $B=512$ samples across feasible target phases. Each marker corresponds to a different target phase $\phi^\star$, with color denoting the number of \emph{unique} successful sequences.

A clear trend emerges: success rate and diversity depend strongly on the target phase. Regions near $240^\circ$–$270^\circ$ exhibit both high success rates and the largest diversity, yielding hundreds of unique realizations that satisfy the efficiency and phase constraints. In contrast, the $0^\circ$–$60^\circ$ region consistently exhibits reduced diversity and lower success rates—precisely the regime where the feasibility envelope indicates intrinsic physical constraints due to the limited tunability and loss characteristics of the stacked JC geometry. To complement these findings, Table~\ref{tab:diversity_fidelity_performance} compares several decoding strategies. As can be deduced, both top-$k$ and top-$p$ sampling maintain high success rates (typically $0.8$ or higher), with top-$k=20$ offering a favorable balance between reliability and structural diversity as it is defined in \eqref{eq:diversity_eq}.

\begin{table}[!t]
\renewcommand{\arraystretch}{1.3}
\caption{Performance of decoding policies within the feasibility envelope.}
\label{tab:diversity_fidelity_performance}
\centering
\begin{threeparttable}
\begin{tabular}{@{}lS[table-format=1.3]S[table-format=1.2]c@{}}
\toprule
{Policy} & {Success Rate\tnote{a}} & {Diversity\tnote{b}} & {Unique\tnote{c}} \\
\midrule
Top-$k$ (10)   & 0.822 & 4.90 & 338 \\
Top-$k$ (20)   & 0.800 & 5.60 & 383 \\
Top-$k$ (50)   & 0.788 & 5.66 & 373 \\
Top-$p$ (0.90) & 0.834 & 5.37 & 378 \\
Top-$p$ (0.95) & 0.818 & 5.47 & 366 \\
\bottomrule
\end{tabular}
\begin{tablenotes}
\small
\item[a] Fraction of generated samples satisfying $|\Delta\phi|\le 5^\circ$ and $|T_{\mathrm{pred}}|^2 \ge 0.95\cdot|T^\star|^2$.
\item[b] Diversity: Median of mean pairwise token $\ell_1$ distance among successful designs.
\item[c] Unique: Median count of distinct successful design sequences.
\end{tablenotes}
\end{threeparttable}
\end{table}

These results highlight that MetaMamba is not restricted to a single deterministic solution. Instead, it can instantly generate \emph{hundreds of distinct high-fidelity unit cell realizations} for a given target. Such diversity is practically valuable, enabling flexibility in fabrication (e.g., feature size constraints) and in secondary objectives (e.g., bandwidth, see also Section~\ref{subsec:broadband}). Crucially, the observed phase dependence demonstrates that the framework attained diversity is correlated with the underlying physical feasibility of MS responses.

\subsection{Broadband Surrogate and Functional Post-Selection}
\label{subsec:broadband}

To further demonstrate the generality and extensibility of the proposed hybrid learning methodology, we extend the calibrated forward surrogate to predict broadband scattering responses over the 18–22~GHz band with $\approx65,000$ frequency responses (FRs). This extension is enabled by the availability of such spectral prediction feature in LAYERS reported in Part I \cite{Marcus2026LAYERS}, which efficiently generates dense frequency responses and thus provides a natural and scalable supervision source for learning dispersive behavior across wide operating bands.

Importantly, the broadband extension does not require redesigning the surrogate architecture or altering the training pipeline. Instead, the forward model is augmented with a set of learned frequency embeddings that condition the shared latent representation on discrete frequency indices. In the present study, the continuous [18,22]~GHz band is discretized with a uniform spacing of 0.1~GHz, yielding $41$ frequency indices. Concretely, after aggregating the layer-wise geometric representation into a compact hidden state, this state is replicated across frequency indices and combined with trainable frequency embeddings, enabling the network to decode frequency-dependent efficiency and phase responses using a common backbone. This mechanism allows the same model to predict either a single operating frequency or a full frequency response by simply adjusting the number of conditioning bins (see Table~\ref{tab:forward_model_details} in Appendix~\ref{app:architectures}), highlighting the modular and adaptable nature of the approach.

The broadband calibration procedure otherwise follows the same hybrid strategy described in Section~\ref{sec:method} and Fig.~\ref{fig:pipeline}. The model is first trained on approximately $65,000$ SA frequency responses generated by LAYERS over the [18,22]\,GHz band (as in Fig.~\ref{fig:pipeline} step (ii), pink block), and subsequently fine-tuned (as in step (iv), pink block) using the \emph{same} set of 1080 CST candidates employed for single-frequency calibration. These CST simulations provide frequency-resolved ground truth across the entire band, allowing broadband correction without repeating the candidate selection process.

Table~\ref{tab:fr_fwd_cst_agreement} compares the predicted and CST-simulated $|T|^2$ and phase responses across the [18,22]~GHz band before and after calibration. While the SA LAYERS predictions capture the general frequency trends, they exhibit noticeable efficiency and phase discrepancies across the band. In contrast, the calibrated broadband forward surrogate reduces mean efficiency errors a factor of 7.2 and phase errors by more than an order of magnitude, demonstrating that the calibration procedure successfully transfers CST-level fidelity across frequencies. 

\begin{figure}[t]
  \centering
  \subfloat[]{%
    \includegraphics[width=0.99\linewidth]{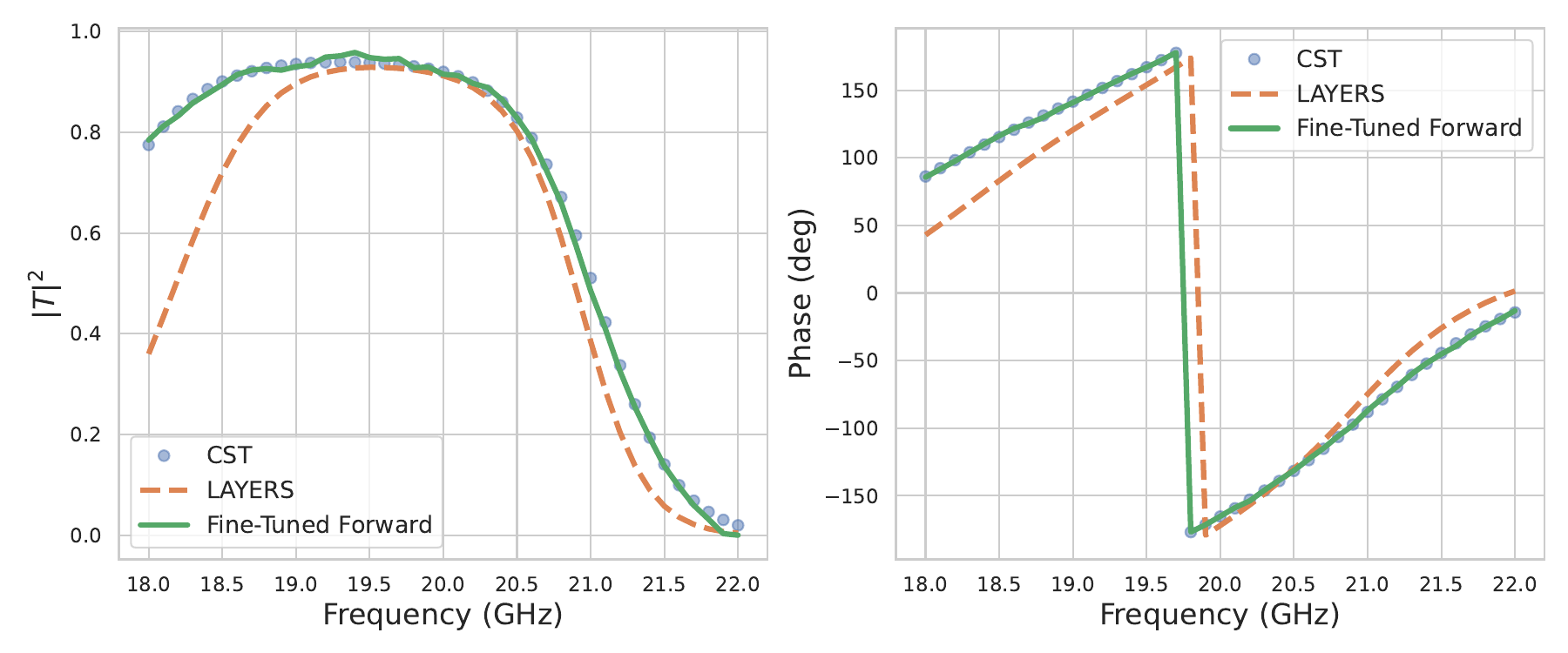}
  }\\[-0.5mm]
  \subfloat[]{%
    \includegraphics[width=0.99\linewidth]{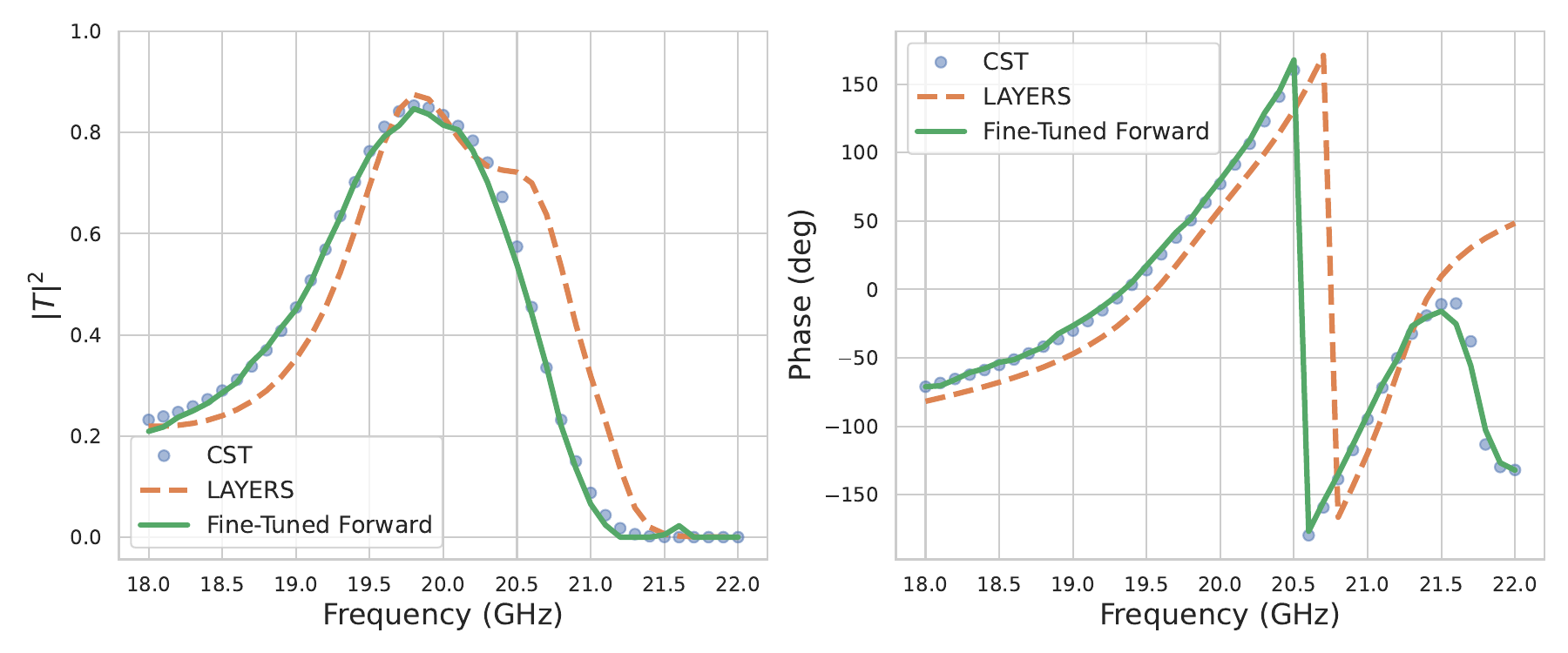}
  }
  \caption{
  Efficiency (left) and phase (right) responses of representative meta-atoms with JC leg lengths (a) $\textbf{W}=(14, 58, 11, 55, 15)$ mil and (b) $\textbf{W}=(22, 17, 15, 20, 24)$ mil over the [18,22]~GHz band, comparing the SA LAYERS predictions (orange dashed lines) with CST ground truth (blue circle markers) and the calibrated broadband forward surrogate (solid green lines).
  }
  \label{fig:fr_calibration}
\end{figure}

Representative spectra for two test-set unit cells are shown in Fig.~\ref{fig:fr_calibration}\textcolor{red}{.} The LAYERS responses (orange curves) follow the overall dispersion trends but deviate systematically from full-wave results, while the calibrated surrogate (green curves) closely overlaps the CST responses (blue curves) across the entire band, with differences that are barely distinguishable at the plotted scale. These results confirm that the broadband surrogate accurately captures both efficiency and phase dispersion of multilayer HMS unit cells.

\begin{figure}[t]
  \centering
  \subfloat[]{%
    \includegraphics[width=0.48\linewidth]{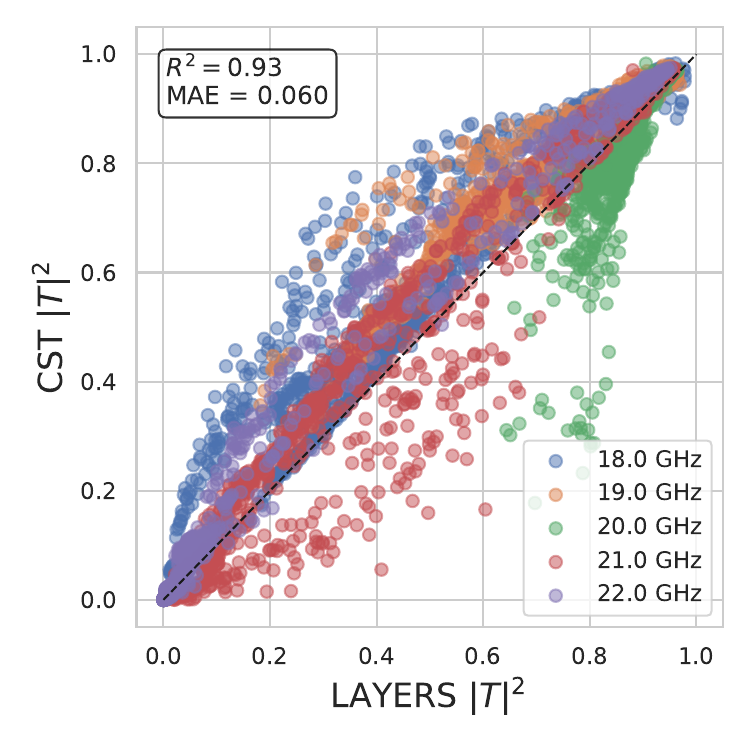}
  }\hfill
  \subfloat[]{%
    \includegraphics[width=0.48\linewidth]{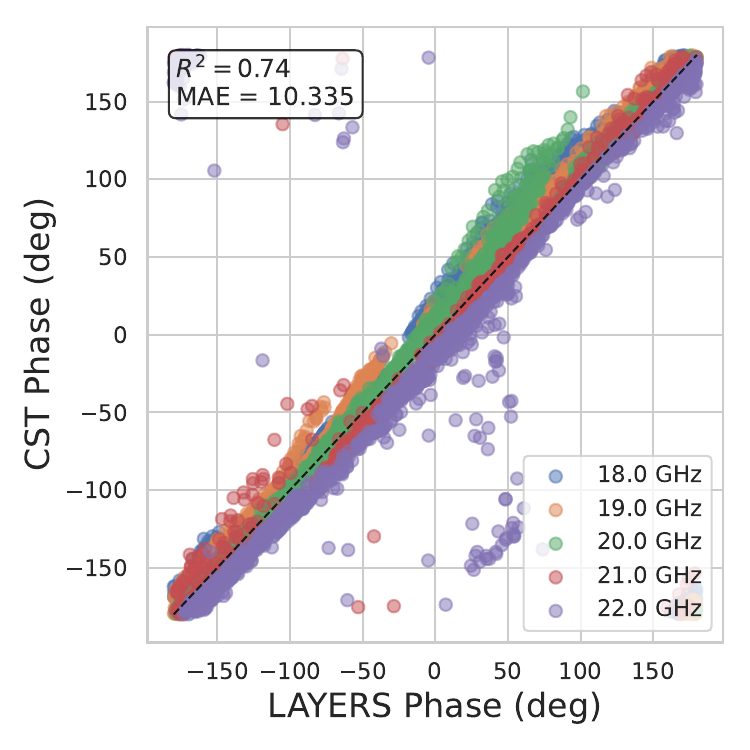}
  }\\[-4mm]
  \subfloat[]{%
    \includegraphics[width=0.48\linewidth]{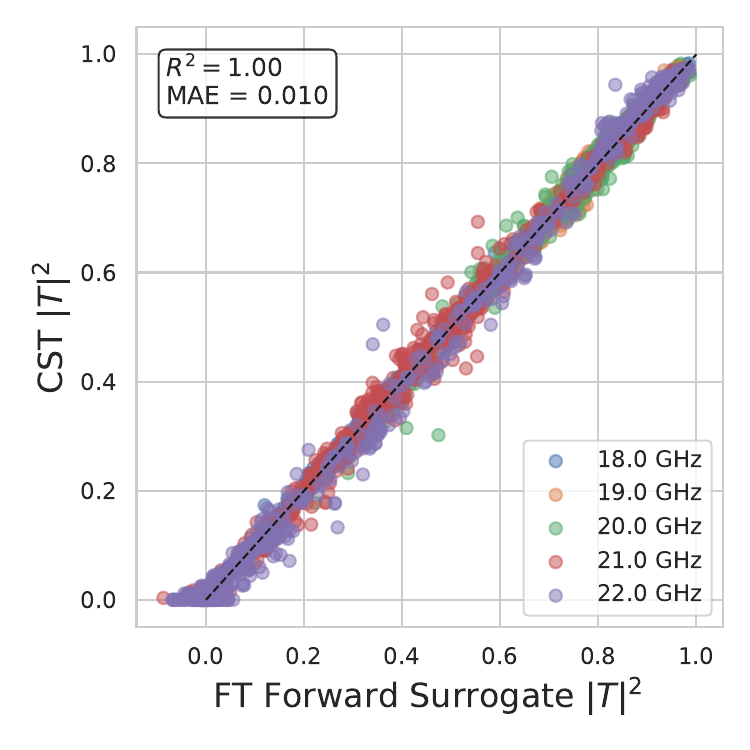}
  }\hfill
  \subfloat[]{%
    \includegraphics[width=0.48\linewidth]{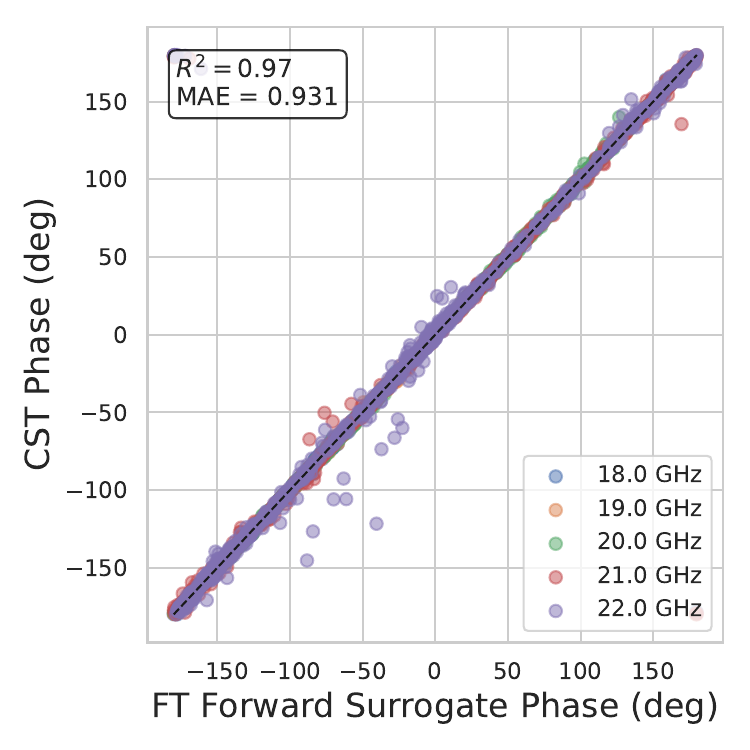}
  }\\[-2mm]
  \caption{Multi-frequency parity plots comparing (a,b) LAYERS and the (c,d) calibrated broadband forward surrogate with CST across the [18,22]\,GHz band. Parity plots of (a,c) transmittance~$|T|^2$ and b,d) phase are shown at five representative frequencies (18, 19, 20, 21, and 22\,GHz).
  }
  \label{fig:fr_parity}
\end{figure}

To complement these representative spectra, Fig.~\ref{fig:fr_parity} presents multi-frequency parity plots comparing the calibrated surrogate with CST across five equally spaced frequencies in the [18,22]~GHz band. These results reveal consistently tight efficiency and phase clustering around the diagonal, confirming that the calibration procedure generalizes across the entire operating band and
not merely at the nominal design frequency. These results confirm that broadband scattering behavior can be captured with the same compact training corpus, underscoring the data efficiency and scalability of the hybrid calibration scheme. From a physical perspective, the model successfully learns the continuous dispersion of the multilayer HMS response.

\begin{table}[t]
\centering
\caption{Broadband agreement with CST on a held-out test set.}

\renewcommand{\arraystretch}{1.1}
\begin{tabular}{lcc}
\hline
 & Efficiency error & Phase error \\
\hline
LAYERS (mean) & 0.0596 & 10.34$^\circ$ \\
FT Forward Surrogate (mean) & 0.0082 & 0.84$^\circ$ \\
\hline
LAYERS (std)  & 0.0827 & 15.64$^\circ$ \\
FT Forward Surrogate (std)  & 0.0137 & 1.83$^\circ$ \\
\hline
LAYERS (min)  & 0.0088 & 2.71$^\circ$ \\
FT Forward Surrogate (min)  & 0.0018 & 0.18$^\circ$ \\
\hline
LAYERS (max)  & 0.1326 & 57.96$^\circ$ \\
FT Forward Surrogate (max)  & 0.0646 & 13.31$^\circ$ \\
\hline
\end{tabular}
\\
\vspace{1mm}
\raggedright
$^a$ Phase errors are wrapped absolute differences in degrees.\\
$^b$ All metrics are computed with respect to the entire [18,22]~GHz band.\\
\label{tab:fr_fwd_cst_agreement}
\end{table}

To further assess the data efficiency of broadband calibration, Table~\ref{tab:fr_calib_ablations} reports an ablation study under varying CST calibration budgets. As the number of full-wave simulations increases from 270 to 1080, both phase and efficiency errors decrease consistently across the band, indicating that additional calibration data primarily refines dispersion accuracy rather than correcting isolated frequency points. Notably, even with 270 CST samples, the surrogate already captures the frequency response with marked precision, while the full 1080-sample budget yields sub-degree mean phase errors and sub-$0.01$ efficiency deviations across the entire operating band.

\begin{table}[t]
\centering
\caption{Broadband calibration ablation of the broadband forward surrogate under varying CST budgets. Errors are evaluated over the [18,22]~GHz band and reported as absolute differences in $|T|^2$ (efficiency) and wrapped phase errors in degrees, evaluated in the same 162-candidate test set.}
\renewcommand{\arraystretch}{1.15}
\begin{tabular}{lcccccc}
\toprule
\textbf{Budget} & \textbf{Metric} & \textbf{Mean} & \textbf{Std} & \textbf{Min} & \textbf{Max} \\
\midrule
270 & Phase (deg) & 1.72 & 5.16 & 0.24 & 23.94 \\
    & Efficiency  & 0.0165 & 0.0265 & 0.0035 & 0.0857 \\
\midrule
540 & Phase (deg) & 1.26 & 2.94 & 0.21 & 15.16 \\
    & Efficiency  & 0.0130 & 0.0209 & 0.0029 & 0.1057 \\
\midrule
1080 & Phase (deg) & 0.84 & 1.83 & 0.18 & 13.31 \\
     & Efficiency  & 0.0082 & 0.0137 & 0.0018 & 0.0646 \\
\bottomrule
\end{tabular}
\label{tab:fr_calib_ablations}
\end{table}

\begin{figure}[t]
  \centering
  \subfloat[]{%
    \includegraphics[width=0.99\linewidth]{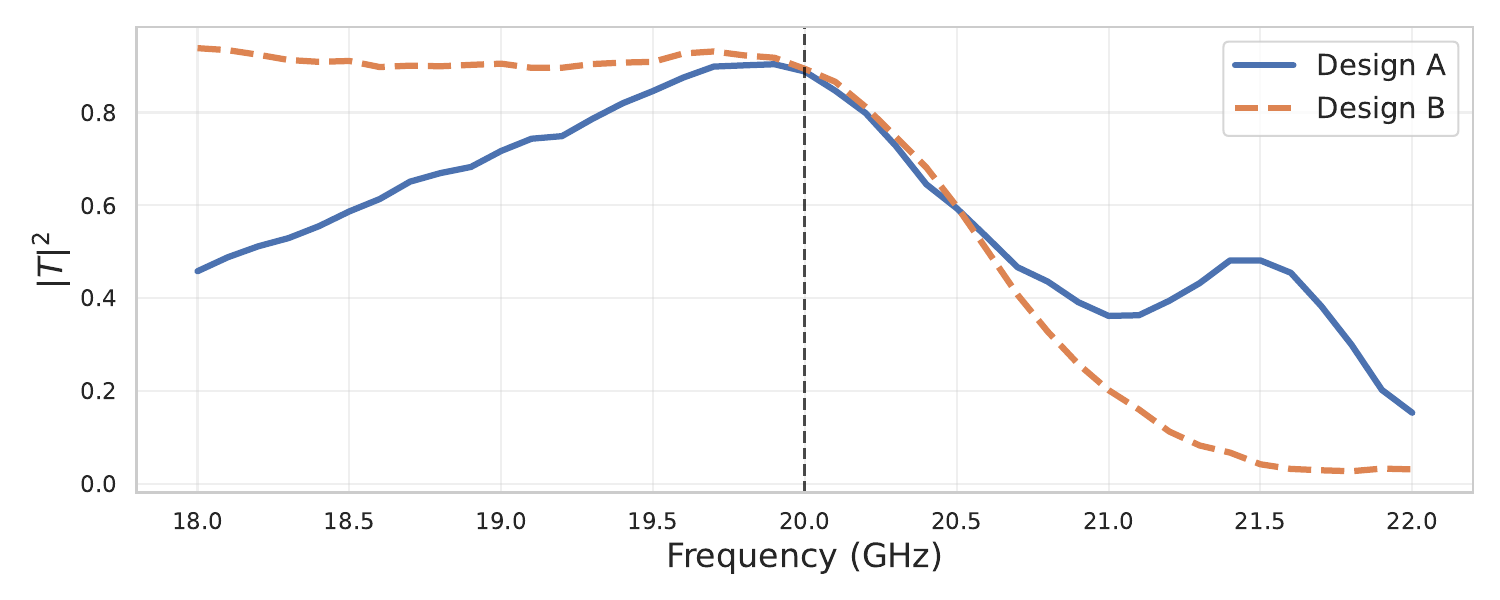}
  }\\[-0.5mm]
  \subfloat[]{%
    \includegraphics[width=0.99\linewidth]{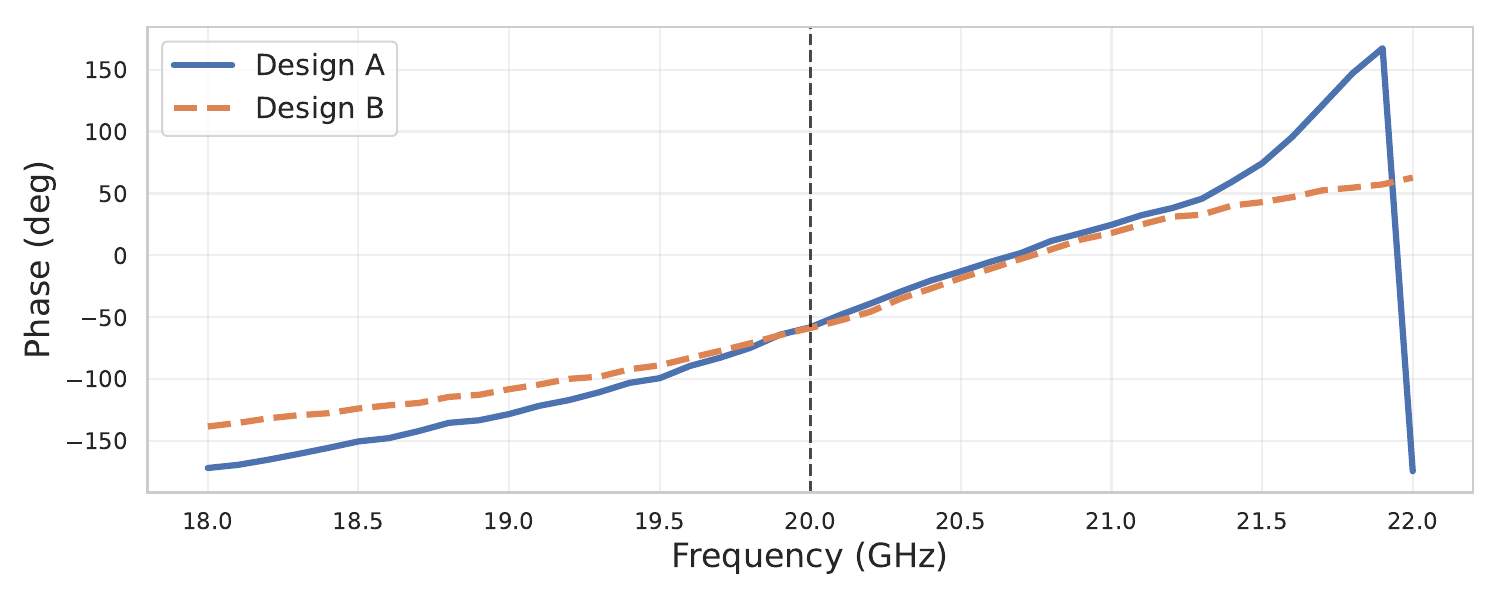}
  }
    \caption{Representative frequency responses of two inverse-generated
    unit cells: (a) efficiency and (b) phase. Both designs math the target scattering response at 20~GHz, yet exhibit distinct behavior across the [18.22]~GHz band.
    The generated sequences of design A (blue solid line) and design B (orange dashed lines) are
    $\mathbf{W}_{A} = (15, 75, 11, 15, 26)$ mil
    and
    $\mathbf{W}_{B} = (6, 5, 17, 8, 5)$ mil, respectively.
    }
  \label{fig:fr_pairs}
\end{figure}

Beyond demonstrating scalability, the broadband surrogate enables assessment of the functional diversity among inverse-generated designs. When the model is used to evaluate multiple unit cells synthesized by the AR-Mamba inverse generator for the same nominal 20~GHz target, it reveals that designs exhibiting nearly identical transmission magnitude and phase at 20 GHz display distinct frequency-dependent characteristics over [18,22]~GHz. An illustration of such representative examples is shown in Fig.~\ref{fig:fr_pairs}, where each frequency response curve corresponds to a unique inverse-generated geometry achieving the same target phase $\phi\approx300^\circ$ and transmission power efficiency $|T|^2\approx0.9$ at 20~GHz, yet showing different off-frequency slopes and bandwidths. Such functional variation is expected, since even when multiple geometries realize the same nominal scattering, the internal mode structure and interlayer coupling differ, leading to different off-resonant behavior.

This capability complements the optimization-based broadband design workflow presented in Part~I ~\cite[Section~III-D]{Marcus2026LAYERS}, where the SA solver is used to define a single unit cell at a time by trading nominal-frequency objectives against band-averaged performance. Here, MetaMamba instead exploits the one-to-many solution set, it instantly generates many nominally compliant geometries and screens their frequency response with the calibrated broadband surrogate.

In practice, this diversity introduces a powerful new design degree of freedom: once a set of cells satisfying the nominal-frequency specification is obtained, the broadband surrogate allows post-selection of those variants exhibiting desirable secondary traits—such as broader bandwidth, flatter group delay, or improved stability near resonance—without retraining the generator or running additional full-wave simulations. Thus, generative diversity supports functional specialization in frequency and subsequently in design tasks.

\subsection{Computational Efficiency and Runtime Analysis}
\label{subsec:runtime}

An important advantage of the proposed hybrid SA--generative framework is its exceptionally favorable computational footprint relative to traditional full-wave–driven workflows. In this context, Table~\ref{tab:runtime} summarizes the sample counts and runtimes for all major stages of the pipeline (Fig.~\ref{fig:pipeline}) in both the single-frequency and broadband regimes, with all measurements obtained using the computational resources detailed in Appendix~\ref{app:resources}. The results reveal a clear and highly desirable trend: \emph{the only stages whose cost grows meaningfully when extending to broadband operation are the data-generation components}—namely, the construction of the SA dataset $\mathcal{D}_{\mathrm{SA}}$ and the compact CST calibration set $\mathcal{D}_{\mathrm{FW}}$. All learning stages, including surrogate pretraining, calibration
fine-tuning, inverse-model training, and AR decoding, remain lightweight and complete within minutes to a few hours on a single GPU.
This division of labor is a direct consequence of the framework’s architecture. The LAYERS model supplies large-scale, low-cost supervision that captures the underlying physics across a wide design space, while the CST calibration set provides only a modest set of high-fidelity
corrections. Once the calibrated surrogate is obtained, both forward prediction and inverse generative design are effectively instantaneous, enabling thousands of unit-cell evaluations or generations in seconds. Notably, even in the broadband case—which traditionally imposes a major computational burden—the overall runtime remains dominated by a one-time CST batch, with learning and inference costs nearly unchanged. These findings highlight the practical scalability of MetaMamba and its ability to deliver CST-level accuracy at a fraction of the simulation expense required by conventional full-wave approaches.

Two additional observations further underscore the practicality of the workflow. First, although we simulated 1080 CST samples for calibration as a conservative upperbound, our ablation analysis (see Table~\ref{tab:calib_ablations}) shows that substantially fewer simulations are sufficient to achieve near--CST-level agreement. This implies that the overall design cycle can be shortened dramatically while retaining high predictive fidelity. In contrast, traditional ML-driven approaches for metasurface inverse design typically require thousands to tens of thousands of high-fidelity full-wave simulations for effective training~\cite{Naseri2021TAP,naseri2022gan,Oliver2022reflect,Niu2023OJAP,Liu2024_gan_fullspace,Chen2025controlnet,Yang2025_tandem,jung2024res}. Second, if one knows in advance that the target application requires accurate broadband behavior, the CST calibration can be performed \emph{only} on frequency response simulations. Because each broadband CST run inherently includes the nominal single-frequency slice, the single-frequency CST simulations become unnecessary, reducing the calibration burden even further. 
The MetaMamba workflow therefore delivers CST-level accuracy at a fraction of the simulation expense required by conventional approaches, offering a scalable and practically deployable tool for multilayer MS design.

\begin{table}[t]
    \centering
    \caption{Sample counts and runtimes for major stages of the MetaMamba workflow. corresponding to the pipeline presented in Fig.~\ref{fig:pipeline}. Hardware specifications appear in Appendix~\ref{app:resources}.}
    \label{tab:runtime}
    \begin{tabular}{lcc}
        \toprule
        \textbf{Stage} & \textbf{Samples} & \textbf{Runtime} \\
        \midrule
        \multicolumn{3}{l}{\textbf{\textit{Single-frequency}}} \\
        SA dataset ($\mathcal{D}_{\mathrm{SA}}$) 
            & $5.24\times10^{5}$ 
            & $2$ h \\
        CST dataset ($\mathcal{D}_{\mathrm{FW}}$) 
            & $1080$ 
            & $\approx2$ days \\[2pt]
            Surrogate pretraining 
            & $5\times10^{5}$ 
            & $1$ h \\
            Surrogate fine-tuning 
            & $270-1080$ 
            & $< 1/2$ h \\
            Forward solving
            & $O(10^{4})$ 
            & 1 sec \\
            Inverse-model training 
            & $2\times10^6$ 
            & $1.5$ h \\
        Generative decoding 
            & $O(10^{4})$ 
            & 1 sec \\
        \midrule
        \multicolumn{3}{l}{\textbf{\textit{Broadband}}} \\
        SA dataset ($\mathcal{D}_{\mathrm{SA}}$) 
            & $65\times10^3$ 
            & $12$ h \\
        CST dataset ($\mathcal{D}_{\mathrm{FW}}$) 
            & $1080$ 
            & $\approx10$ days \\[2pt]
        Surrogate pretraining 
            & $64\times10^3$ 
            & $1/2$ h \\
        Surrogate fine-tuning 
            & $270-1080$ 
            & $< 1/2$ h \\
        Forward solving
            & $O(10^{4})$ 
            & 1 sec \\
        \bottomrule
    \end{tabular}
\end{table}

\section{Conclusion}

This work introduced MetaMamba, a sequence-aware generative pipeline for the inverse design of transmissive unit cells. By combining the SA scheme introduced in Part~I, a CST-calibrated Bi-Mamba forward surrogate, and an AR-Mamba inverse generator, the framework enables CST-validated generation of diverse five-layer unit cells that achieve field transmission magnitudes $|T|>0.9$ across the full $0$–$2\pi$ phase range, while maintaining power transmission efficiencies above $90\%$ over approximately $72\%$ of the phase span, relying on as few as 270 CST-labeled calibration samples to reach near--CST-level agreement on a held-out full-wave test set.

Several insights emerge from this study. First, with the foundation laid by the SA scheme, forward model calibration requires only a modest CST budget. The ablation results in Tables~\ref{tab:calib_ablations} and ~\ref{tab:fr_calib_ablations}, show that even much smaller calibration sets than our initial assumption yield near substantial accuracy, highlighting the data efficiency of the approach. Second, the inverse generator instantly produces multiple distinct high-fidelity designs for each target, enabling exploitation of the one-to-many nature of inverse mapping. Third, framing unit cell synthesis as a sequential generation task opens a natural path toward scaling to deeper stacks, richer geometries, and other complex electromagnetic objectives. Finally, the integration with the LAYERS SA framework developed in Part~I demonstrates the broader vision of this two-part study: combining fast physics based modeling with modern generative sequence learning to enable scalable, data-efficient inverse design pipelines for multilayer MSs.

Looking ahead, the MetaMamba framework readily supports several impactful extensions without requiring fundamental architectural changes. In particular, the conditioning sequence can be enriched to encode broader electromagnetic specifications, including reflection characteristics, broadband frequency responses, incidence-angle dependence, and polarization diversity. Overall, our results demonstrate that data-efficient, generative inverse design for electromagnetic MSs is not only feasible but can match full-wave simulation accuracy at a fraction of the computational cost.

\appendices




\section{Architectures}\label{app:architectures}

For completeness and reproducibility, this appendix details the exact network architectures used for the forward (Bi-Mamba) surrogate and the inverse (AR-Mamba) generator, as well as the common Mamba-2 building block shared by both models. All quantitative results reported in the Section~\ref{sec:results} are obtained using the architectures specified herein. While Section~\ref{sec:method} motivates these choices at a conceptual level, the present appendix provides implementation-level detail, including dimensionalities, parameter counts, and data flow, enabling faithful reimplementation and informed architectural modification.

The forward and inverse networks share a common state space backbone but differ in how sequence information is processed and emitted.
The forward surrogate addresses a deterministic regression task, mapping a fixed-length layer sequence to a global scattering response, and therefore employs bidirectional processing to capture bidirectional interlayer coupling.
In contrast, the inverse generator addresses an intrinsically ill-posed, one-to-many synthesis task, and is formulated as a causal AR model that emits layer parameters sequentially.
Table~\ref{tab:architecture_comparison} summarizes these high-level architectural distinctions, while subsequent Table~\ref{tab:forward_model_details} and Table~\ref{tab:inverse_model_details} provide detailed specifications for each model.

\begin{table}[t]
\caption{Forward vs. inverse Mamba architectures (summary).}
\label{tab:architecture_comparison}
\centering
\footnotesize
\begin{tabularx}{\columnwidth}{@{}lYY@{}}
\toprule
\makecell[l]{Aspect} & \makecell[l]{Forward model} & \makecell[l]{Inverse model} \\
\midrule
Primary architecture & \makecell[l]{Bi-Mamba \\ + MLP heads} & \makecell[l]{Causal Mamba \\ (LM head)} \\
SSM variant & Mamba-2 & Mamba-2 \\
$d_{\text{model}}$ & 256 & 256 \\
Intermediate dim. & 512 & 512 \\
Layers & 6 & 6 \\
SSM state $d_{\text{state}}$ & 64 & 64 \\
Parameters & 5.22M & 5.00M \\
Input length & 5 & variable (AR) \\
Output length & 3 & variable (AR) \\
Input type & \makecell[l]{Continuous lengths $W$} & \makecell[l]{Continuous $S$ and \\ discrete $W$ tokens} \\
Output type & Continuous $S$ & Discrete $W_{1:5}$ \\
Normalization & RMSNorm & RMSNorm \\
Fused AddNorm & yes & yes \\
Objective & \makecell[l]{Regression (MSE)} & \makecell[l]{teacher forced CE} \\
\bottomrule
\end{tabularx}
\end{table}

From a signal-flow perspective, the forward model first embeds the layer-wise geometric parameters into a shared latent space, processes the resulting sequence through a Bi-Mamba backbone, and pools the sequence dimension before projecting to scattering outputs, as depicted in Fig.~\ref{fig:Bi-Mamba}. In broadband operation, this pooled geometric state is replicated across discrete conditioning indices and augmented with learned conditioning embeddings (indexed by frequency in the present study), enabling frequency-dependent decoding using a shared backbone. The inverse model follows the opposite mapping: a compact embedding of the target response initializes the sequence, after which discrete layer parameters are generated token-by-token through a causal Mamba backbone and a language-model head (Fig.~\ref{fig:ARMamba}). Despite these differences, both models rely on the same underlying state space mechanism to propagate context efficiently across the sequence, either bidirectionally or causally.

\begin{table}[t]
\caption{Forward model additional details.}
\label{tab:forward_model_details}
\centering
\footnotesize
\begin{tabularx}{\columnwidth}{@{}lY@{}}
\toprule
Component & Specification \\
\midrule
Input projection & Linear: $1 \to 256$ \\
Mamba backbone & $6\times$ Mamba-2 layers (bidirectional) \\
Bi-directional processing & Forward and reversed sequence passes with additive hidden-state fusion \\
Pooling (Sequence Dimension) & Linear: $5 \to 1$ \\
Frequency indices (FI) & \makecell[l]{1 (single frequency 20~GHz);\\ 41 (broadband [18,22]~GHz)} \\
Learned conditioning embeddings & $|FI| \times 256$ , added to pooled state\\
Magnitude head & MLP: $256 \to 512 \to 1\cdot |FI|$ \\
Phase head & MLP: $256 \to 512 \to 2 \cdot |FI|$ \\
Output & Concatenate phase \& magnitude \\
Activation & SiLU \\
Parameter count & 5.22M \\
\bottomrule
\end{tabularx}
\end{table}

\begin{table}[t]
\caption{Inverse (AR-Mamba) model additional details.}
\label{tab:inverse_model_details}
\centering
\footnotesize
\begin{tabularx}{\columnwidth}{@{}lY@{}}
\toprule
Component & Specification \\
\midrule
Mixed embeddings & \makecell[l]{Float: $1 \to 256$ \\ Discrete: vocab ($81$) $ \to 256$} \\
Mamba backbone & $6\times$ Mamba-2 layers (causal) \\
Causality & AR generation of $W_{1:5}$ \\
LM head & Linear: $256 \to 81$ \\
Float seq. length & 3 ($\sin\phi,\cos\phi, |T|^2$) \\
Vocabulary & 81 tokens (per $W_n$) \\
Parameter count & 5.00M \\
Decoding & top-$k$/ top-$p$ \\
\bottomrule
\end{tabularx}
\end{table}

Both the forward and inverse networks are composed of identical Mamba-2 blocks, whose internal structure is summarized in Table~\ref{tab:mamba_block_structure}. This block combines local convolutional mixing with a state space module, enabling efficient modeling of both short-range and long-range dependencies within the layer sequence.

\begin{table}[t]
\caption{Mamba-2 block structure (used in both models).}
\label{tab:mamba_block_structure}
\centering
\footnotesize
\begin{tabularx}{\columnwidth}{@{}l l l Y@{}}
\toprule
Component & Op. & Dim. & Notes \\
\midrule
Input & $x$ & $(B,L,256)$ & Input sequence \\
Pre-norm & RMSNorm$(x)$ & $(B,L,256)$ & Input normalization \\
Proj-in & Linear & $256 \to 512$ & Expand $d_{model}$. \\
Activation & SiLU & $(B,L,512)$ & Nonlinearity \\
Conv1D & Conv1D $(k{=}4)$ & $(B,L,512)$ & Local context mixing \\
SSM & Mamba-2 & $(B,L,512)$ & State space ($d_{\text{state}}{=}64$) \\
Gate & Element-wise $\times$ & $(B,L,512)$ & Gated activation \\
Proj-out & Linear & $512 \to 256$ & Back to model dim. \\
Residual & $x +$ proj-out & $(B,L,256)$ & Skip connection \\
\bottomrule
\end{tabularx}
\end{table}

\section{Training Hyperparameters}
\label{app:hyperparams_config}

This appendix summarizes the key hyperparameters and training schedules used for the Bi-Mamba surrogate and AR-Mamba inverse generator. These settings complement the architectural specifications in the Appendix~\ref{app:architectures}.

The training configuration reflects the distinct roles of the two models. The forward surrogate is trained to minimize a smooth regression objective and therefore relies solely on validation loss for learning-rate scheduling and early stopping. The inverse generator, in contrast, is optimized exclusively using the cross-entropy loss, while its downstream physical fidelity is assessed using a reconstruction error metric. The latter is used only for model evaluation and as an auxiliary early-stopping criterion, and does not enter the optimization objective. These choices are reflected in the hyperparameter settings summarized below.

\begin{table}[!t]
\centering
\scriptsize
\caption{Training hyperparameters.}
\label{tab:training_hparams}
\begin{tabular}{lcc}
\hline
\textbf{Parameter} & \textbf{Bi-Mamba} & \textbf{AR-Mamba} \\
\hline
Batch size & 1024 & 1024 \\
Optimizer & AdamW & AdamW \\
Weight decay & 0.01 & 0.01 \\
Dropout & 0.0 & 0.0 \\
Init. LR & 0.001 & 0.001 \\
Scheduler & ROP & ROP \\
LR Patience & 2 & 2 \\
LR reduction factor & 0.1 & 0.1 \\
$L_{\text{rec}}$ stall patience & --- & 5 \\
Term. LR & $1{\times}10^{-5}$ & $1{\times}10^{-5}$ \\
Early stopping condition & LR $<$ Term. LR & LR $<$ Term. LR \textbf{or} $L_{\text{rec}}$ stall \\
\hline
\end{tabular}
\end{table}

\noindent
Both models were trained using the AdamW optimizer (weight decay $10^{-2}$), which provided stable convergence across both regression and AR objectives. A reduce-on-plateau learning-rate schedule was employed, whereby the learning rate was reduced by a factor of $0.1$ if the validation loss failed to improve for a fixed number of epochs (patience), as specified in Table~\ref{tab:training_hparams}. Early stopping criteria were used to terminate training when further improvements became negligible, thereby preventing overfitting and avoiding unnecessary computational cost.

\section{Fine-Tuning Hyperparameters}
\label{app:ft_hparams}

This appendix reports the hyperparameters used for fine-tuning the surrogate model on CST calibration data while retaining knowledge from the pre-trained SA surrogate (Table~\ref{tab:ft_hparams}).
The calibration strategy employs several key design choices. First, a \emph{rehearsal schedule} interleaves CST and SA batches at a 2:1 ratio, with CST batches of 128 samples and SA batches of 1024 samples, ensuring the model sees high-fidelity corrections while maintaining broad coverage. Second, \emph{differential learning rates} are applied: the pretrained backbone uses $1\times10^4$ while the prediction heads use $5\times10^4$, allowing faster adaptation of task-specific layers while preserving learned representations. Third, two hyperparameters control the balance between high- and low-fidelity data:
\begin{itemize}
  \item \emph{Rehearsal loss weight} (Eq.~\ref{eq:calib}): $\lambda_{FW}=1.0$ and $\lambda_{SA}=0.2$ mix full CST loss with 20\% scaled SA loss for gradient computation.
  \item \emph{Validation weighting}: $\alpha=0.7$ prioritizes CST validation performance when selecting model checkpoints.
\end{itemize}
Training proceeds for 100 epochs using a cosine scheduler with warmup. This configuration provides stable calibration, retaining SA generalization while aligning closely with CST ground truth and effectively mitigating catastrophic forgetting.

\begin{table}[!t]
\centering
\scriptsize
\caption{Fine-tuning hyperparameters for surrogate calibration.}
\label{tab:ft_hparams}
\begin{tabular}{l c}
\hline
\textbf{Parameter} & \textbf{Value} \\
\hline
Epochs & 100 \\
Batch size (CST, high-fidelity) & 128 \\
Batch size (SA, rehearsal) & 1024 \\
Rehearsal rate (CST:SA) & 2:1  \\
Backbone learning rate & $1.0 \times 10^{-4}$ \\
Head learning rate & $5.0 \times 10^{-4}$ \\
Backbone weight decay & $1.0 \times 10^{-2}$ \\
Head weight decay & $1.0 \times 10^{-4}$ \\
Dropout & 0.0 \\
Scheduler & Cosine with warmup \\
Rehearsal loss weight $\lambda_{FW}$; $\lambda_{SA}$ & 1.0; 0.2 \\
Validation weighting $\alpha_{FW}; \alpha_{SA}$ & 70\%; 30\% \\
\hline
\end{tabular}
\end{table}

\section{Computation Resources}
\label{app:resources}

This appendix details the hardware and software environments used for data generation, surrogate calibration, and inverse design experiments.

\subsection{Model Training and Inference}
All surrogate and inverse model training, fine-tuning, and inference were performed on a single NVIDIA L40S GPU. 
The software environment was:
\begin{itemize}
    \item Operating system: Linux 6.8.0-62-generic (x86\_64, glibc 2.39)
    \item Python: CPython 3.10.16
    \item CUDA: 12.8
    \item PyTorch: 2.2
\end{itemize}

\subsection{SA Data Generation}
The SA datasets were generated on a standard PC with:
\begin{itemize}
    \item Operating system: Windows 11 Pro
    \item Processor: Intel Core i7-1355U (10 cores, 12 threads)
    \item RAM Memory: 16 GB
\end{itemize}

\subsection{CST Simulations}
All full-wave simulations were performed using \emph{CST Microwave Studio 2023} on a dedicated workstation at the CommLab facility, with the following specifications:
\begin{itemize}
    \item Workstation: HP Z8 G4 (model Z3Z16AV)
    \item Operating system: Microsoft Windows 11 Pro for Workstations
    \item CPUs: 2$\times$ Intel Xeon Gold 6244 @ 3.60~GHz (8 cores each, 16 logical processors per CPU)
    \item Installed memory: 192~GB RAM
    \item GPU: NVIDIA Quadro GV100 (32~GB)
\end{itemize}
The frequency-domain solver was used, which does not support GPU acceleration. Accordingly, all CST runtimes reported in Table~\ref{tab:runtime} correspond to CPU-only execution.

\section*{Acknowledgment}
The authors thank Doron Klepach from FVMat for fruitful discussions.

\balance
\bibliographystyle{IEEEtran}
\bibliography{tap_references}

\end{document}